\documentclass[psfig,12pt]{article}
\usepackage{epsfig}
\begin{document}
\def\be{\begin{equation}}
\def\bea{\begin{eqnarray}}
\def\ee{\end{equation}}
\def\eea{\end{eqnarray}}
\def\Tr{\rm Tr}
\def\d{\partial}
\def\la{\lambda}
\def\eps{\epsilon}
\def\zb{{\bar z}}
\def\wb{{\bar w}}
\def\vb{{\bar v}}
\def\db{{\bar \d}}
\def\ie{{\it i.e.}}
\def\eg{{\it e.g.}}
\def\cf{{\it c.f.}}
\def\cN{{\cal N}}
\def\etal{{\it et.al.}}
\def\etc{{\it etc.}}
\def\half{{\frac{1}{2}}}
\def\pa{\partial}
\def\CR{\nonumber\\}
\def\rt{\rightarrow}
\def\a{\alpha}
\def\t{\theta}
\def\IR{\relax{\rm I\kern-.18em R}}
\begin{titlepage}
\begin{flushright}
TAUP-TH 2913/10\\
WIS/07/10-JUN-DPPA\\
\end{flushright}
\bigskip
\def\thefootnote{\fnsymbol{footnote}}

\begin{center}
{\Large {\bf Holographic MQCD }} \\
\end{center}

\bigskip

\begin{center}

{\large Ofer Aharony$^a$\footnote{Ofer.Aharony@weizmann.ac.il},
David Kutasov$^{b}$\footnote{dkutasov@uchicago.edu}, Oleg Lunin$^c$\footnote{lunin.7@gmail.com},
}
\end{center}

\begin{center}

{\large Jacob Sonnenschein$^{d,e}$\footnote{cobi@post.tau.ac.il}, Shimon Yankielowicz$^{d,e}$\footnote{ shimonya@post.tau.ac.il}}

\end{center}

\begin{center}
\textit{{$^a$\it Department of Particle Physics and Astrophysics,
\\ Weizmann Institute of Science, Rehovot 76100, Israel}}
\end{center}

\begin{center}
\textit{{$^b$\it  EFI and Department of Physics, University of Chicago,
Chicago, IL 60637, USA}    }
\end{center}
\begin{center}
\textit{{$^c$\it
Department of Physics and Astronomy, University of Kentucky,}
\centerline{Lexington, KY 40506, USA
}}
\end{center}
\begin{center}
\textit{{$^d$\it School of Physics and Astronomy,\\
The Raymond and Beverly Sackler Faculty of Exact Sciences,\\
Tel Aviv University, Ramat Aviv, 69978, Israel}}
\end{center}
\begin{center}
\textit{{$^e$\it Albert Einstein Minerva Center, Weizmann Institute of Science, Rehovot 76100, Israel}}
\end{center}

\vfil

\renewcommand{\thefootnote}{\arabic{footnote}}

\noindent
\begin{center}
{\bf Abstract}
\end{center}

We study a brane configuration of $D4$-branes and $NS5$-branes
in weakly coupled type IIA string theory, which describes in a particular
limit  $d=4$ ${\cal N}=1$ $SU(N+p)$ supersymmetric QCD with $2N$
flavors and a quartic superpotential. We describe the geometric realization
of the supersymmetric vacuum structure of this gauge theory. We
focus on the confining vacua of the gauge theory, whose holographic
description is given by
the MQCD brane configuration in the near-horizon geometry of $N$
$D4$-branes. This description,  which gives an embedding of MQCD
into a field theory decoupled from gravity,  is valid for $1\ll p \ll N$,
in the limit of large five dimensional `t Hooft couplings for the color and flavor groups.
We analyze various properties of the theory in this limit, such as the spectrum
of mesons, the finite temperature behavior, and the quark-anti-quark potential.
We also discuss the same brane configuration on a circle, where it gives a
geometric description of the moduli space of the Klebanov-Strassler cascading
theory, and some non-supersymmetric generalizations.



\vfill

\begin{flushleft}
{\today}
\end{flushleft}

\end{titlepage}

\section{Introduction and summary}

It is believed that in the limit of a large number of colors (\eg\ large $N_c$ for an
$SU(N_c)$ gauge theory), many gauge theories can be reformulated as weakly coupled closed
string theories with $g_s \sim 1/N_c$, following the ideas of  \cite{'tHooft:1973jz}. This reformulation may facilitate
the understanding of non-perturbative properties like confinement and chiral symmetry
breaking. For most gauge theories the (higher dimensional) target space of the dual string
theory, which is usually referred to as the bulk, is highly curved; so far we do not have good
quantitative methods to analyze such string theories. For some specific gauge theories, the
dual string theory lives in a weakly curved space.  In these cases, the dynamics of the
strongly coupled large $N_c$ gauge theory,  which lives on a space isomorphic to the boundary
of the bulk space, can be studied in detail (see \cite{AharonyTI} for a review).

Adding $N_f$ flavors in the fundamental representation of an $SU(N_c)$ gauge group corresponds
in the dual string theory to adding $N_f$ $D$-branes in the bulk. For small $N_f$ ($N_f \ll N_c$),
the theory with flavor can be studied by adding the dynamics of open strings ending on
these $D$-branes; \eg, when the gauge theory has a global $SU(N_f)$ symmetry, the
symmetry currents correspond in the bulk to gauge fields on the $D$-branes. When
$N_f$ becomes of the same order as $N_c$, the open string coupling on the $D$-branes,
$g_sN_f \sim N_f/N_c$, is not small, and there is no reason to believe that a weakly
coupled string description exists.\footnote{For a recent review of some approaches to
holography for theories with $N_f \sim N_c$, and further references, see \cite{Nunez:2010sf}.}

One way to obtain a weakly coupled string theory for $N_f\sim N_c$ is by gauging the flavor
group.  In this case the usual arguments of the 't Hooft limit imply that there should be a weakly
coupled dual string theory, but it is generally not the same as the original theory, unless the
gauge coupling of the flavor group is weak. Of course,
when the flavor group is weakly gauged we generally do not expect to get a weakly curved string
theory dual, but such a dual may exist when the flavor group is strongly coupled.

In this paper we analyze an example of a large $N_c$ gauge theory with the flavor group gauged,
for which a weakly curved string dual exists when the flavor group is strongly coupled.
Our large $N_c$ gauge theory will be four dimensional, but we will gauge the flavor group by
coupling it to five dimensional gauge fields (with a UV completion given by a six dimensional
conformal field theory). The five dimensional flavor gauge theory is IR-free; our weakly curved
gravity dual is useful when the interesting physics happens at energies at which this theory is
strongly coupled, but there is also a different limit of the same theory where the flavor gauge
theory is weakly coupled and the physics is that of the original four dimensional
gauge theory with $N_f\sim N_c$ flavors.

Our field theory arises as a decoupling limit of the brane configuration shown in figure
\ref{origin} below. This configuration (see \eg\ \cite{GiveonSR} for a review of the dynamics of
this and related brane configurations) involves
$N$ $D4$-branes which intersect (along $3+1$ dimensions) two $NS5$-branes with different
orientations, and a stack of $p$ additional $D4$-branes which stretch between the fivebranes.
This brane system preserves $d=4$ $\cN=1$ supersymmetry and gives, in a certain decoupling
limit, $d=4$ $\cN=1$ $SU(N+p)$ supersymmetric QCD (SQCD) with $2N$ flavors of ``quarks''
in the fundamental representation of the gauge group. For vanishing superpotential, this theory
flows in the IR to a non-trivial fixed point \cite{Seiberg:1994pq}. The brane
construction gives rise to a quartic superpotential which preserves an $SU(N)\times SU(N)$ flavor
symmetry. The resulting gauge theory has multiple vacua, some of which are confining (in other vacua,
some of the gauge symmetry is spontaneously broken).

We will discuss a different decoupling limit, in which the flavor symmetry is gauged by coupling
it to five dimensional gauge fields (which are the gauge fields on the semi-infinite $D4$-branes
in figure \ref{origin}). We will argue that when $p \ll N$, and the five dimensional gauge theory
is strongly coupled (at the characteristic energy scale of the four dimensional dynamics), this
theory has a simple string dual. In particular, the confining vacua can be described by placing
the MQCD \cite{Witten:1997ep} fivebrane in the near-horizon geometry of $N$ $D4$-branes.\footnote{Holography
for the $D4$-brane geometry was developed in \cite{ItzhakiDD}.} This gives us a controllable background which
is continuously related (by changing parameters) to SQCD, similar to the way that the backgrounds
of \cite{Polchinski:2000uf,ks,Maldacena:2000yy} are continuously related to the pure $d=4$ $\cN=1$
supersymmetric Yang-Mills (SYM) theory. In our case, there are additional fields living in a higher
dimensional space, that only decouple in the limit that their gauge coupling goes to zero.
Our main purpose in this paper is to investigate the properties of the resulting system.

We begin in section \ref{setups} by describing the brane configuration and its different limits.
For $N=0$, our
brane configuration is similar to MQCD, which is obtained by taking the string coupling in figure
\ref{origin} to be large. We study the brane system in a different limit, where the type IIA string coupling
is small, and the MQCD fivebrane is an $NS5$-brane carrying fourbrane flux. To study the  dynamics
of the fivebrane semiclassically, we take the five dimensional 't Hooft coupling of the $SU(p)$ gauge
theory on the $D4$-branes to be large.

For large $N$, the confining vacua of the gauge theory are described  by embedding the MQCD
brane configuration (for gauge group $SU(p)$) into the near-horizon geometry of $N$ $D4$-branes. In this
sense our discussion provides an embedding of MQCD into a field theory which is decoupled from
gravity (the decoupling limit from gravity of the MQCD brane configuration itself is still unknown).
Unlike the original MQCD configuration, this enables us to have normalizable states (corresponding to
four dimensional particles) coming from the
MQCD brane, and to study the theory at finite temperature.
We discuss both the brane configuration corresponding to the SQCD theory discussed above, and
its generalization to the case where the $D4$-branes live on a circle; in the latter case our field theories
are $SU(N)\times SU(N+p)$ gauge theories similar to the ones that appear in the Klebanov-Strassler
cascade \cite{ks}, but we study these theories in a different range of parameters from \cite{ks}. As in
the non-compact case, we find that some features of the cascading theories (like their moduli space)
are realized in our limit as well, while other features are different.

In section \ref{fieldth} we analyze the field theories corresponding to our brane configurations,
and in particular study their moduli space and match it to the dual string description,
finding precise agreement whenever the string theory computation is under control. In particular, we
find an elegant geometrical description for the complicated moduli space \cite{Dymarsky:2005xt} of
the Klebanov-Strassler cascading theory.
In section \ref{Tfotbaths} we compute the spectrum of operators and states in our string theory
dual. We find that the spectrum is generally continuous from the four dimensional point of view,
because some higher dimensional fields do not decouple in our limit, but there are also some discrete
states which may be continuously connected to the mesons of SQCD.

In section \ref{QCDstring} we discuss some of the energy scales in our problem, and in particular
the quark-anti-quark potential. We show that for some range of parameters this is dominated by the
five dimensional IR-free physics, but that there is also a range of parameters (and of quark-anti-quark
distances) for which it is dominated by the four dimensional confining physics. In section \ref{finiteT}
we discuss the behavior of our system at finite temperature, showing that at all finite temperatures
the confining phase has a higher free energy than the Higgs phase, in agreement with field theory
expectations. Finally, in section \ref{nonsusygen}
we discuss some non-supersymmetric generalizations of our construction, including a case where there is
a first order finite temperature phase transition (at which the fivebrane falls into the horizon). Clearly
there are many possible generalizations of our setup, both supersymmetric (\eg\ theories related
to $d=3$ $\cN=2$ SQCD) and non-supersymmetric; we leave their analysis to future work.

\section{The brane construction}\label{setups}

\subsection{ $\cN=1$ Supersymmetric Yang-Mills from type IIA string theory}

\begin{figure}[]
\begin{center}
\includegraphics[width= 90mm]{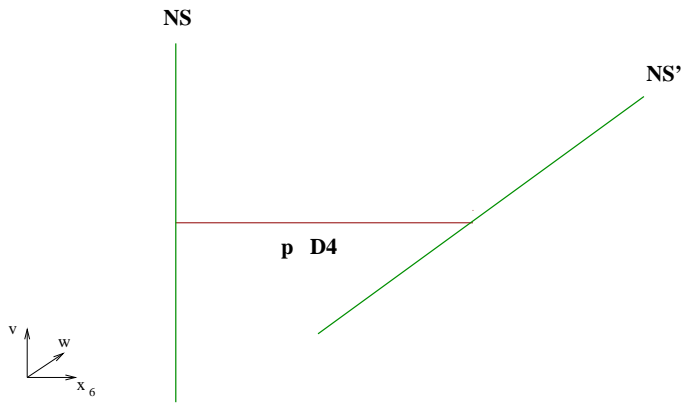}\end{center}
\caption{The brane system realizing $d=4$ $\cN=1$ SYM with gauge group $U(p)$.\label{puresym}}
\end{figure}

Pure $d=4$ $\cN=1$ SYM with gauge group $U(p)$ can be realized in type IIA string theory
as the low energy limit of the system of intersecting $D4$-branes and $NS5$-branes
depicted in figure \ref{puresym}. All the branes in the figure are extended in the $\IR^{3,1}$
labeled by $(x_0, x_1, x_2, x_3)$. The fivebranes are further extended in
\bea\label{Thebranes}
&NS:\qquad v=x_4+ix_5, \nonumber\\
&NS': \qquad w=x_8+ix_9,
\eea
while the $p$ fourbranes form a line segment of length $L$ in the $x_6$ direction,
\bea\label{ldfour}
-{L\over2}\le x_6\le {L\over2}.
\eea
As reviewed in \cite{GiveonSR}, this brane configuration preserves $\cN=1$ supersymmetry
in the $3+1$ dimensions common to all the branes, $(0123)$. All the fields in the $D4$-brane
gauge theory other than the $U(p)$ gauge fields and gauginos get masses of order $1/L$ due
to the boundary conditions at $x_6=\pm L/2$.

The classical four dimensional gauge coupling is given by
\bea\label{gaugefour}
g_{YM}^2={g_sl_s\over L}.
\eea
Later, we will be interested in the large $p$ limit of this system, in which the strength of interactions
is governed by the `t Hooft coupling,
\bea\label{lambdafour}
\lambda^{(4)}_p=g_{YM}^2 p={\lambda_p\over L};\qquad \lambda_p \equiv g_sl_sp.
\eea
$\lambda_p$ (which has units of length) is the $4+1$ dimensional `t Hooft coupling of the
$D4$-brane theory.

Quantum mechanically, the coupling of the four dimensional gauge theory runs with the scale.
One can view $1/L$ as the UV cutoff of this theory, and $\lambda^{(4)}_p$ as the value of the
coupling at the UV cutoff scale. In order for the dynamics of the brane configuration to reduce
to that of $\cN=1$ SYM at energies well below the cutoff scale, this coupling must be taken
to be very small, $\lambda^{(4)}_p\ll 1$.

The classical picture of $D4$-branes ending on $NS5$-branes (figure \ref{puresym}) is
qualitatively modified by $g_s$ effects  \cite{Witten:1997sc}. To exhibit these effects, it is
convenient to view type IIA string theory at finite $g_s$ as M-theory compactified on a
circle of radius $R=g_sl_s$. Both the $D4$-branes and the $NS5$-branes correspond from the
eleven dimensional point of view to $M5$-branes, either wrapping  the M-theory circle ($D4$-branes),
or localized on it ($NS5$-branes). The configuration of figure \ref{puresym}
lifts to a single $M5$-brane, which wraps $\IR^{3,1}$ and a two dimensional surface in the
$\IR^5\times S^1$ labeled by $(v,w,z)$. Here
\bea\label{defz}
z\equiv x_6+ix_{11},
\eea
and $x_{11}$ parameterizes the M-theory circle, $x_{11}\sim x_{11}+2\pi R$. The shape of
the fivebrane in $\IR^5\times S^1$ is described by the equations \cite{Witten:1997ep}
\bea\label{WittProf}
vw=\xi^2 ,\qquad v=\xi e^{-z/pR}=\xi e^{-z/\lambda_p},
\eea
where without loss of generality we can choose $\xi$ to be real and positive.
Note that:
\begin{itemize}
\item{The classical brane configuration of figure \ref{puresym} lies on the surface $vw=0$,
while the quantum shape is deformed away from this surface. In particular, the $D4$-branes,
that are classically at $v=w=0$, are replaced in the quantum theory by a tube of width $\sim\xi$
connecting the (deformed) $NS$ and $NS'$-branes. Indeed,  defining  the radial coordinate
$u$ via
\bea\label{defuvw}
u^2=|v|^2 + |w|^2,
\eea
we see that (\ref{WittProf}) satisfies $u\ge \sqrt{2}\xi$. One can interpret $\xi/l_s^2$ as the
brane analog of the dynamically generated  scale of the $\cN=1$ SYM theory. }
\item{Classically, the brane configuration is confined to the interval (\ref{ldfour}), while quantum
mechanically it extends to arbitrarily large $|x_6|$. For example, for large positive $x_6$, the
fivebrane takes the shape
\bea\label{asshape}
z\simeq \lambda_p\ln(w/\xi);\;\;\; v\simeq 0.
\eea
One can think of (\ref{asshape}) as describing an $NS'$-brane deformed by the $D4$-branes
ending on it from the left. Since these $D4$-branes are codimension two objects on the fivebrane,
they give rise to a deformation of it that does not go to zero at infinity. }
\item{As usual, isometries in the bulk give rise to global symmetries of the field theory on the
branes. Consider the following three $U(1)$ symmetries: $U(1)_{45}$, corresponding to rotations in
the $(45)$ plane, $U(1)_{89}$, corresponding to rotations in the $(89)$ plane, and $U(1)_{11}$, corresponding to translations in $x_{11}$. Classically, the first two symmetries are preserved by
the brane configuration of figure \ref{puresym}, while the third one is broken by the positions of the
$NS5$-branes. Quantum mechanically, the asymptotic $x_6\to\infty$ shape (\ref{asshape}) breaks
one linear combination of $U(1)_{89}$ and $U(1)_{11}$, while its analog as $x_6 \to -\infty$ breaks a
linear combination of $U(1)_{45}$ and $U(1)_{11}$. The full brane configuration (\ref{WittProf})
preserves  a single $U(1)$ symmetry, whose action is given by
\bea\label{branesym}
v \to e^{i\alpha} v, \qquad
w \to e^{-i\alpha} w, \qquad x_{11} \to x_{11} - \alpha \lambda_p.
\eea
Note that the other two $U(1)$
symmetries are broken by the asymptotic boundary conditions, so they are broken explicitly from
the point of view of the field theory living on the branes. As discussed in \cite{Witten:1997ep},
since pure ${\cal N}=1$ SYM theory has no unbroken $U(1)$ global symmetry, all states that are
charged under the symmetry (\ref{branesym}) are expected to decouple
in any limit that leaves only the degrees of freedom of this four dimensional gauge theory.  One
of the broken $U(1)$ symmetries discussed above is actually not completely broken by the
asymptotic shape (\ref{asshape}) and its $x_6\to-\infty$ analog. It is broken to $Z_p$, which is
further spontaneously broken by the full brane configuration (\ref{WittProf}). This $U(1)$ symmetry
can be identified with the (anomalous) R-symmetry of the four dimensional SYM theory.}
\item{At large $u$, the distance between the $NS$ and $NS'$-branes goes to infinity; hence
the separation $L$ appears to be ill defined. This is not surprising, since  (\ref{gaugefour})
relates $L$ to the four dimensional gauge coupling, which changes with the scale.
To define it, one can take the radial coordinate $u$  (\ref{defuvw}) to be bounded, $u\le u_\infty$,
and demand that at $u=u_\infty$, $x_6=\pm L/2$. Assuming that the four dimensional
`t Hooft coupling $\lambda^{(4)}_p$  (\ref{lambdafour}) is very small, this gives the following
relation among the different scales:
\bea\label{lcutoff}
\xi=u_\infty \exp\left(-L/2\lambda_p\right)= u_\infty \exp\left(-1/2\lambda^{(4)}_p\right).
\eea
This is the brane analog of the relation between the QCD scale and the gauge coupling
in SYM theory, with $u_\infty/l_s^2$ playing the role of a UV cutoff.\footnote{The corresponding
energy scale is in general different from the KK scale $1/L$ mentioned above.} As in SYM, we can
remove the cutoff, by sending $u_\infty, L/\lambda_p\to\infty$ while keeping the ``QCD scale''
$\xi/l_s^2$ fixed.
}
\item{Although the curved fivebrane is non-compact, and in particular extends to infinity in $\IR^5$,
the non-trivial dynamics is restricted to the intersection region. The only low energy modes that live
on the fivebrane at large $(v,w,x_6)$ are $5+1$ dimensional free fields that describe the position of
a single fivebrane, and their superpartners. One can take a limit in which these fields
decouple, and only the $3+1$ dimensional physics remains, but we will not do that here.}
\end{itemize}

As one varies the parameters of the brane system of figure  \ref{puresym}, the language in terms
of which its IR dynamics is most usefully described changes. The pure SYM description is valid
when $\lambda^{(4)}_p$ (\ref{lambdafour}) and the dynamically generated scale $\xi$ (\ref{WittProf})
are small. In this regime the brane description reduces to the gauge theory one.

In other regions in parameter space the dynamics can
be studied by analyzing the geometry of the branes. One such region is obtained by sending
$g_s\to\infty$, \ie\ taking the radius of the M-theory circle, $R$, to be much larger than the
eleven dimensional Planck length. If $\xi$ is also taken to be sufficiently large, (\ref{WittProf})
describes a large and smooth $M5$-brane, whose dynamics can be studied semiclassically. The low
energy theory of this fivebrane is known as MQCD; it was discussed in \cite{Witten:1997ep} and
subsequent work. Since it is related to $\cN=1$ SYM by a continuous
deformation of the parameters of the brane configuration, and no phase transitions are expected
along this deformation, the two theories are believed to be in the same universality class. However,
many of their detailed features are expected to be different.

In this paper we will focus on a different region in the parameter space of the brane system.
We will take the type IIA string coupling to be small, but $p$ to be large, such that the five
dimensional `t Hooft coupling is large, $\lambda_p\gg l_s$. As mentioned above, for large $L$
and small $\xi$ the low energy dynamics of this system reduces to $\cN=1$ SYM. On the other hand,
if $\xi$ is sufficiently large, the fivebrane described by (\ref{WittProf}) is weakly curved (in string
units). Thus, the situation is similar to that in MQCD, except for the fact that the string coupling
is weak. This implies that the fivebrane in question is an $NS5$-brane, which also carries RR
six-form flux ($D4$-brane charge).

The shape of this $NS5$-brane is given by the reduction of the eleven dimensional profile
(\ref{WittProf}) to ten dimensions. It is parameterized by two functions of $x_6$, $u$ and
$\alpha$, which are defined (on the fivebrane) by
\bea\label{SpecEqn}
v=ue^{i\phi}\cos(\a),\qquad w=ue^{-i\phi}\sin(\a).
\eea
Looking back at (\ref{WittProf}), we see that $u$ and $\alpha$ are given by
\bea\label{NSUsol}
u=\xi\sqrt{2\cosh \left(\frac{2x_6}{\lambda_p} \right)},\qquad \tan(\a)=\exp\left(2x_6/\lambda_p\right).
\eea
The configuration  (\ref{WittProf}) also involves a non-zero expectation value for the (compact)
scalar field which labels the position of the $NS5$-brane  along the M-theory circle. As $\phi$ in
(\ref{SpecEqn}) varies between $0$ and $2\pi$, the scalar field winds $p$ times
around the circle, giving the $NS5$-brane its $D4$-brane charge.

When $\lambda_p,\xi\gg l_s$, the shape (\ref{NSUsol}) is weakly curved and a semiclassical
description should be reliable. This description depends on whether the back-reaction of the
branes on the geometry can be neglected. The $p$ $D4$-branes modify the geometry around
them significantly up to distances of order $(\lambda_p l_s^2)^{1/3}$. Thus, if $\xi$ satisfies the
constraint
\bea\label{xiconst}
\xi^3\gg \lambda_p l_s^2,
\eea
one can neglect the back-reaction and treat the curved fivebrane as a probe in flat
spacetime. Otherwise, the back-reaction is important, and one can try to replace the
branes by their geometry, and describe their low-energy dynamics using holography. This is an
interesting problem that we will leave to future work.

The regime (\ref{xiconst}), where the curved fivebrane (\ref{SpecEqn}), (\ref{NSUsol}) can be treated
as a probe, is quite analogous to MQCD. The validity of the semiclassical analysis of the fivebrane relies
in this case on the `t Hooft large $p$ limit rather than on large $g_s$, as in \cite{Witten:1997ep}, but many
of the qualitative properties are similar.

\subsection{A holographic embedding of MQCD}
\label{holemb}

\begin{figure}[t]
\begin{center}
\includegraphics[width= 100mm]{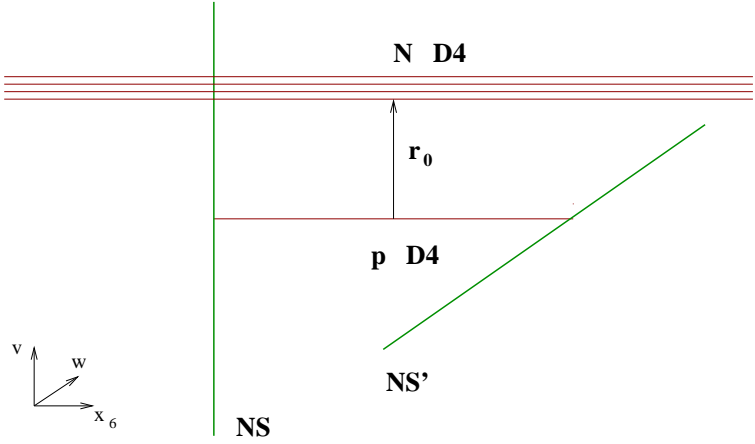}\end{center}
\caption{The brane system with $N$ additional infinite $D4$-branes.\label{massive}}
\end{figure}

We next embed the brane configuration of figure \ref{puresym} in a larger system, which is more amenable to
a holographic analysis. To do this, we add $N$ infinite $D4$-branes stretched in the $x_6$ direction (see
figure \ref{massive}). These branes do not
break any of the supersymmetries preserved by the configuration of figure \ref{puresym}. Thus, we
can place them anywhere in the $\IR^5$ labeled by $(v,w, x_7)$, without influencing the shape of
the curved $NS5$-brane (\ref{NSUsol}). This can be seen directly by replacing the $N$ $D4$-branes by
their geometry, and studying the dynamics of the curved $NS5$-brane (\ref{NSUsol}) in that geometry.
This description should be valid for $N\gg p$, and we will restrict to this (``probe'') regime below.

Viewing the $N$ $D4$-branes as $M5$-branes wrapped around the M-theory circle,
their eleven dimensional geometry is given by\footnote{It is well known that dimensional
reduction of this geometry gives the correct description of $D4$-branes in type IIA string
theory. Thus, the discussion below is valid in the weakly coupled type IIA limit as well.}
\bea\label{TheMetric}
ds^2&=&H^{-1/3}\left(dx_\mu^2+dx_6^2+dx_{11}^2\right)+
H^{2/3}\left(|dv|^2+|dw|^2+dx_7^2\right),\nonumber\\
C_6&=&H^{-1}d^4x\wedge dx_6\wedge dx_{11},\qquad
H=1+\frac{\pi\lambda_N l_s^2}{|\vec r-\vec r_0|^3}\,,
\eea
where  $\lambda_N=g_sl_sN$ is the $4+1$ dimensional `t Hooft coupling of the $N$
$D4$-branes (defined as in (\ref{lambdafour})),
$\mu=0,1,2,3$, $\vec r=(v,w, x_7)$ labels position in $\IR^5$, with $\vec r=0$ corresponding to
the classical position of the $p$ $D4$-branes in figure \ref{massive}, and $\vec r_0\in \IR^5$ is the
position of the $N$ $D4$-branes.

For $g_s\not=0$, the $NS5$-branes and $p$ $D4$-branes form a
curved $M5$-brane whose shape may be obtained by plugging the ansatz
\bea\label{nsansatz}
v=u(x_6)e^{i\phi(x_{11})}\cos(\a(x_6)),\qquad w=u(x_6)e^{-i\phi(x_{11})}\sin(\a(x_6)),
\eea
into the  fivebrane worldvolume action. Parametrizing the $M5$-brane worldvolume by the
coordinates $(x_{\mu}, x_6, x_{11})$, the induced metric corresponding to (\ref{nsansatz}) takes
the form
\bea
ds_{ind}^2=H^{-1/3}\left\{dx_\mu^2+\left[1+H\left((u\a')^2+(u')^2\right)\right]dx_6^2+
\left(1+H(u{\dot \phi})^2\right)dx_{11}^2\right\},
\eea
where $u'\equiv \pa_{x_6} u$, $\a' \equiv \pa_{x_6} \a$ and $\dot\phi\equiv \pa_{x_{11}}\phi$.
The Lagrangian is\footnote{Here and below, we omit the tension of the fivebrane, which
appears as a multiplicative factor in the Lagrangian.}
\bea\label{Laguncomp}
L=H^{-1}\sqrt{1+H(u{\dot \phi})^2}\sqrt{1+H((u\a')^2+(u')^2)}-H^{-1}.
\eea
The equations of motion imply that ${\dot\phi}$ must be constant; thus
$\phi=x_{11}/pR=x_{11}/\lambda_p$.
The Noether charges $J, E$ associated with the invariances under the shifts of $\alpha$ and
$x_6$, respectively, are then given by
\bea\label{ccc}
J &=&\frac{u^2\a'\sqrt{1+ H u^2/\lambda_p^2}}{\sqrt{1+H\left[(u\a')^2+(u')^2\right]}},\CR
 E&=& H^{-1}-\frac{H^{-1}\sqrt{1+H u^2/\lambda_p^2}}{\sqrt{1+H\left[(u\a')^2+(u')^2\right]}}.
\eea
Supersymmetric configurations have $E=0$. Substituting into (\ref{ccc}) we find:
\bea\label{ueqzeroe}
\a'&=&J/u^2,\CR
(u')^2&=&\frac{u^2}{\lambda_p^2}-\frac{J^2}{u^2}.
\eea
Note that the equations (\ref{ueqzeroe}) that determine the shape of the supersymmetric $M5$-brane
are independent of the form of the harmonic function $H$, and in particular of the
positions of the $N$ $D4$-branes in figure \ref{massive}. This agrees with the
expectation that there is no force between the various branes. As a check, one
can verify that the profile (\ref{NSUsol}) indeed solves (\ref{ueqzeroe}), with
\bea\label{XiVsJ}
J\lambda_p=2\xi^2.
\eea
In this solution $\a$ goes from $\a(x_6\to -\infty) = 0$ to $\a(x_6\to \infty) = \pi/2$.

So far, our discussion took place in the full type IIA string theory. We next take a
decoupling limit, by omitting the $1$ in the harmonic function $H$ (\ref{TheMetric}).
This corresponds to studying the brane configuration of figure \ref{massive} in the
$(2,0)$ theory of $N$ $M5$-branes, compactified on a circle of radius $R$. One
can think of the curved $M5$-brane (\ref{NSUsol}) as a localized defect in this theory.
The dynamics of the modes of the $(2,0)$ theory contributes to the interactions among
the fields localized on the defect, which include four dimensional $\cN=1$ gauge
superfields. We will see later that  the low energy theory contains some additional higher
dimensional modes, and is thus not purely $\cN=1$ SYM. However, it does not contain any
gravitational or stringy dynamics, in contrast to MQCD and its weakly coupled
analog described in the previous subsection.

\begin{figure}[t]
\begin{center}
\includegraphics[width= 100mm]{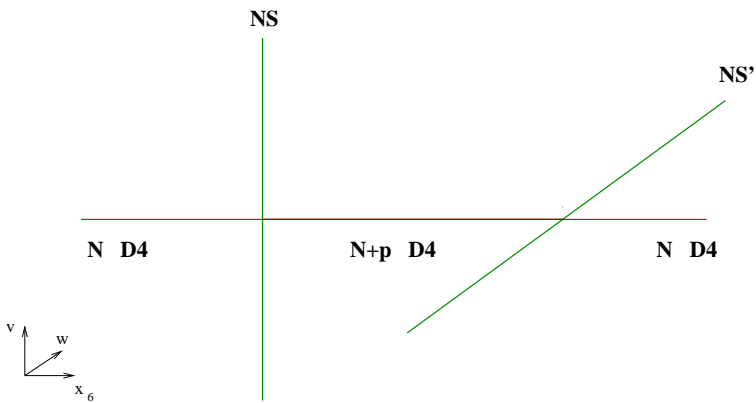}\end{center}
\caption{The brane system of figure \ref{massive} at the origin of its moduli space.\label{origin}}
\end{figure}

As mentioned above, the shape of the curved $M5$-brane (\ref{NSUsol}) is independent of the
position of the $N$ $D4$-branes, $\vec r_0$. Consider the limit $\vec r_0\to 0$, in which all
the $D4$-branes in the classical configuration of figure \ref{massive} are coincident (see figure
\ref{origin}). As is clear from the figure, there are states (corresponding to fundamental strings
stretched between the two stacks of $D4$-branes), whose masses go to zero in this limit. We
will discuss their field theoretic interpretation in the next section; here we note that for
$\vec r_0=0$ the system has some additional supersymmetric vacua.

In the classical brane description, these vacua can be obtained as follows. One or more of the fourbranes
stretched between the $NS5$-branes can connect to semi-infinite branes attached from the left to the
$NS$-brane, and make semi-infinite $D4$-branes stretching from the $NS'$-brane to $x_6=-\infty$.
As is clear from figure \ref{origin}, this leaves behind semi-infinite fourbranes stretching from
the $NS$-brane to $x_6=+\infty$. The semi-infinite branes in the two stacks can now be independently
displaced along the appropriate $NS5$-brane, thus giving rise to a new branch of the moduli space of brane
configurations. In this branch, the number of $D4$-branes stretching between the $NS5$-branes is reduced.

\begin{figure}[t]
\begin{center}
\includegraphics[width= 100mm]{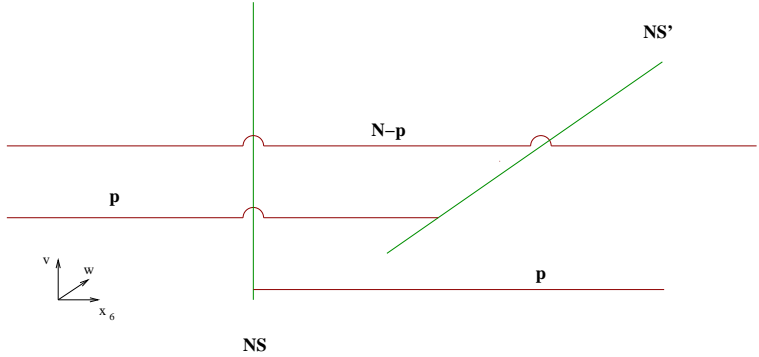}\end{center}
\caption{The ``disconnected'' branch of the moduli space.\label{fullhiggs}}
\end{figure}

When all $p$ $D4$-branes reconnect as described above, the fivebrane splits into two separate
fivebranes, each of which corresponds to a solution of (\ref{ueqzeroe}) with $J=0$. The angle
$\alpha$ (\ref{SpecEqn}) takes the constant values $0$ and $\pi/2$ for the $NS$ and $NS'$-brane,
respectively. The solution to (\ref{ueqzeroe}) is
\be\label{discU}
u = K \exp\left(\pm x_6/\lambda_p\right).
\ee
The number of $D4$-branes stretched all the way from $x_6=-\infty$ to $+\infty$ (\ie\ the
background $D4$-brane flux) decreases to $N-p$ (see figure \ref{fullhiggs}). We will discuss
the field theory interpretation of these extra branches of the moduli space of brane configurations
in the next section.

Before introducing the $N$ extra $D4$-branes, we found that the parameter $\xi$, (\ref{WittProf}), (\ref{NSUsol}),
must satisfy the bound (\ref{xiconst}) in order for the back-reaction of the $p$ $D4$-branes on the geometry
to be negligible. In the near-horizon geometry of the $N$ $D4$-branes, the back-reaction of the $p$ $D4$-branes
is always a subleading effect in $p/N$ (which we are assuming is small). There is still a constraint
coming from the fact that if $\xi$ is small, the tube connecting the $NS5$-branes is located in the
region where the curvature of the near-horizon geometry is large and cannot be trusted. For
$\vec r_0=0$ this constraint takes the form
\bea\label{xinearh}
{\xi\over l_s^2}\gg {1\over\lambda_N}.
\eea
It is much less stringent than (\ref{xiconst}); we will impose it below.
Recall that $\xi/l_s^2$ is the QCD scale associated with the tube (\ref{WittProf});
it is held fixed in the decoupling limit from gravity described above. The constraint (\ref{xinearh})
is precisely the requirement that the $4+1$-dimensional $U(N)$ gauge theory on the $D4$-branes
is strongly coupled at the scale $\xi/l_s^2$ \cite{ItzhakiDD}.

\subsection{Compact $x_6$}\label{compcirc}

\begin{figure}[htp]
\begin{center}
\includegraphics[width= 80mm]{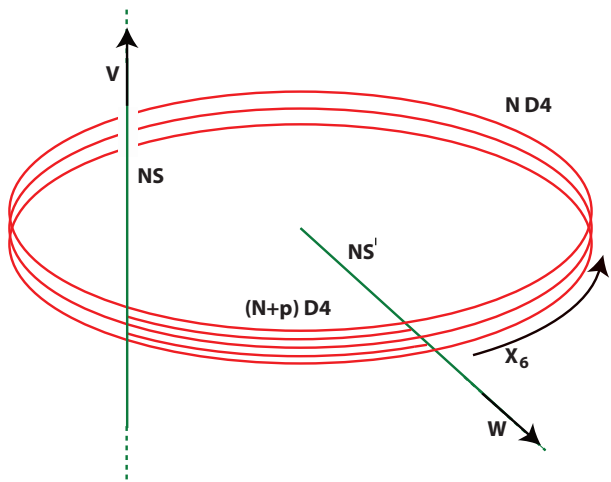}\end{center}
\caption{$N$ $D4$-branes that wrap the $x_6$ circle, and $p$
that are stretched between the $NS$ and $NS'$-branes.\label{compactified}}
\end{figure}

Another interesting brane system is obtained from that of figure \ref{massive} by compactifying the
$x_6$ direction on a circle, $x_6\sim x_6+2\pi R_6$. The resulting configuration (with $\vec r_0=0$)
is depicted in figure \ref{compactified}.
While the bosons living on the branes must satisfy periodic boundary conditions around
the circle, fermions can be either periodic or antiperiodic. Most of our discussion below
will involve the periodic case, in which supersymmetry is unbroken by the boundary
conditions. In this case, the string background is still given by (\ref{TheMetric}),
and the shape of the curved fivebrane by (\ref{SpecEqn}), (\ref{NSUsol}), (\ref{discU}),
with $x_6$ periodically identified.

\begin{figure}[htp]
\begin{center}
\includegraphics[width= 60mm]{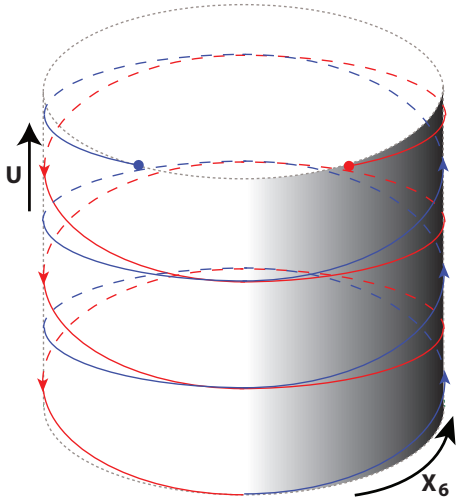}\end{center}
\caption{ For compact $x_6$,  the curved $NS5$-brane spirals down the $(x_6, u)$ cylinder
and then climbs back up. The downward (upward) part of the spiral is colored red (blue).
\label{cascading}}
\end{figure}

On the cylinder labeled by $(x_6,u)$, the fivebrane described by (\ref{SpecEqn}), (\ref{NSUsol})
takes the form depicted in figure \ref{cascading}. Starting at the boundary $u=u_\infty$, it
spirals down the cylinder and then climbs back up to the boundary. In the process, the angle
$\alpha$ changes by $\pi/2$. The disconnected configuration (\ref{discU}) gives rise to a solution
with constant $\alpha$, which spirals down the cylinder and does not climb back.

Although in the covering space of the cylinder the shape of the fivebrane is the same as in the
non-compact case, there are two important differences between the two cases. The first has to do with
the fact that the curved fivebrane carries $p$ units of fourbrane charge. In the non-compact case,
if one fixes $x_6$ and increases the radial coordinate $u$, the flux increases by $p$ units, from
$N$ to $N+p$, when one crosses the fivebrane. For finite $R_6$, increasing $u$
at fixed $x_6$ one encounters the fivebrane multiple times, as it spirals around the cylinder.
At each encounter, the fourbrane flux increases by $p$ units. If we fix the flux to be $N+p$ at the
UV boundary $u=u_\infty$, as we spiral down, it may\footnote{The change of flux between
$u\sim \xi$ and $u_\infty$ is $\Delta N\sim L/R_6$, where $L$ is the distance between the
$NS5$-branes (in the covering space) evaluated at $u_\infty$ (see the discussion around (\ref{lcutoff})). It could be
large or small compared to $N$.}  eventually become comparable to $p$, where the probe
approximation breaks down.

The second difference between the compact and non-compact cases concerns the vacuum structure of the model.
In the non-compact case, fivebranes described by (\ref{SpecEqn}), (\ref{NSUsol}) with different
values of $\xi$ correspond to different theories, much like SYM theories with different values
of $\Lambda_{QCD}$. Indeed, for large $u$ one has (see (\ref{asshape}))
\bea\label{nonnorm}
x_6\sim \lambda_p\ln \left({u\over\xi}\right),
\eea
and changing $\xi$ corresponds to a non-normalizable mode in the geometry (\ref{TheMetric}).
In the compact case, profiles with
\bea\label{xinn}
\xi=\xi_n=\xi_0 e^{2\pi n R_6\over\lambda_p},\qquad n\in Z
\eea
give rise to the same asymptotics (\ref{nonnorm}), and thus describe different vacua of a single theory.
The integer $n$ is bounded from above by the requirement $\xi_n<u_\infty$ and from below by the
breakdown of the probe approximation described above. Later we will discuss the field theory
interpretation of these vacua.

When the radius $R_6$ becomes too small, the adjacent coils in figure \ref{cascading} approach
each other, and one can no longer ignore their back reaction on the geometry. The requirement that
this does not happen leads to the inequality
\bea
R_6\gg \left(\frac{\la_p^4}{\la_N}\right)^{1/3}.
\eea
Another constraint on $R_6$ comes from requiring that upon reducing the geometry (\ref{TheMetric})
to IIA string theory, the radius of the $x_6$ circle at $u\sim\xi$ should be much larger than the string
length. This leads to the constraint
\bea
R_6\gg \left(\frac{\la_N l_s^2}{(\xi/l_s^2)^3}\right)^{1/6}.
\eea

The brane configuration of figure \ref{compactified} is T-dual to the system studied
in \cite{ks}, which involves regular and fractional $D3$-branes placed at a conifold
singularity in type IIB string theory. The Klebanov-Strassler theory describes the limit
of our configuration as $R_6 \to 0$ and $\lambda_p, \lambda_N \to 0$, keeping the four
dimensional QCD scale $\xi/l_s^2$ fixed; this is an opposite limit to the limit of strong
five dimensional gauge coupling that we focus on in this paper. On the type IIB side the brane sources are
replaced by varying three-form and five-form fluxes. In particular, the gauge-invariant
charge associated with $F_5$
changes along the throat; this change can be attributed to a sequence of Seiberg
dualities (see section 3). One can also define a Page charge associated with the
five-form \cite{Benini:2007gx}, which is conserved along the flow, but is not gauge-invariant
(though its value modulo $p$ is gauge-invariant).

The type IIA story is simpler. In the probe approximation ($p\ll N$) discussed in this paper,
the four-form field strength has explicit sources (the spiraling fivebrane), and there is only
one charge, which is both gauge-invariant and conserved. It changes as one moves in
the radial direction due to the presence of sources.

The system discussed in \cite{ks} is known to have a rich vacuum
structure, some of which is described by regular type IIB supergravity solutions (in a certain regime in
the parameter space of brane configurations). The type IIA description
in terms of a fivebrane with $D4$-brane charge winding around the cylinder in figure \ref{cascading} is
valid in a different regime in parameter space, but the (supersymmetric) vacuum structure is expected to
be the same. We will comment on this comparison further below.

\section{Field theory}\label{fieldth}

In this section we will discuss the field theory that governs the low energy dynamics
of the brane configurations of figures \ref{origin} and \ref{compactified}, and its relation
to that of the brane systems described above.

\subsection{Non-compact $x_6$}

We start with the brane configuration of figure \ref{origin}, and consider its low energy dynamics in
the weak coupling limit. The low energy effective field theory on the $D4$-branes contains two types
of excitations: four dimensional fields localized in the intersection region, and five dimensional
fields living on the semi-infinite fourbranes at $|x_6|\ge L/2$.

The four dimensional fields include a $U(N+p)$ $\cN=1$ SYM theory, and two sets of $N$
(anti) fundamental chiral superfields $(Q_L^i,\tilde Q_{L,i})$, $(Q_R^\alpha,\tilde Q_{R,\alpha})$;
$i,\alpha=1,2,\cdots, N$. The latter are coupled by the superpotential \cite{AharonyJU}
\be
\label{suppot}{W_\lambda=\lambda (\tilde Q_L\cdot Q_R)(\tilde Q_R\cdot Q_L)}~,
\ee
where the scalar product stands for contraction of the $U(N+p)$ color indices, and the flavor
indices $(i,\alpha)$ are contracted between the two gauge-invariant bilinears. This
theory is invariant under a $U(N)_L\times U(N)_R$ global symmetry, which acts on the
indices $(i,\alpha)$ (note that this is a non-chiral symmetry). This symmetry is the
global part of the gauge group of the
semi-infinite $D4$-branes on the left and right  of figure \ref{origin}.

The theory is also invariant under a $U(1)$ global symmetry under which $Q_L$,
$\tilde Q_L$ have charge one, and $Q_R$, $\tilde Q_R$ have charge  minus one.
This symmetry corresponds in the brane picture to the $U(1)$ symmetry (\ref{branesym}).
To see that, it is useful to think about it as a difference of two R-symmetries, acting
on $(Q_L, \tilde Q_L)$ and $(Q_R,\tilde Q_R)$, respectively. These R-symmetries
correspond in the brane language to rotations in the $w$ and $v$ planes. Hence, their
difference, which does not act on the supercharges, acts on $v$ and $w$ as in (\ref{branesym}).

While the chiral superfields are localized in $x_6$ (at $x_6=\pm L/2$), the vector
superfields are five dimensional fields living on the line segment (\ref{ldfour}).
At energies of order $1/L$, one starts seeing the massive Kaluza-Klein (KK) states,
both of the gauge field and of the other $D4$-brane modes (the five transverse scalars
and fermions). The four dimensional gauge coupling is given in terms of the five
dimensional one by an analog of equation (\ref{lambdafour}), $\lambda^{(4)}_{N+p}=\lambda_{N+p}/L$.
In this section we assume that it is small at the KK scale, though this is not true when the
gravitational approximation of the holographic description of the previous section is valid.

In addition to the fields mentioned above, the brane system contains five dimensional fields
living on the two stacks of semi-infinite fourbranes in figure \ref{origin}.
They are described by a $4+1$ dimensional SYM theory with sixteen supercharges, broken down
to eight supercharges by the boundary conditions at $x_6=\pm L/2$. The supersymmetry is
further broken down to $\cN=1$ by the couplings of the five dimensional fields to the four
dimensional ones. These couplings can be read off from figure \ref{origin} by examining the
effects of geometric deformations. In particular, the superpotential (\ref{suppot}) receives an
additional contribution of the form
\be
\label{extrasuppot}{W_\phi=\lambda_\phi\left(\tilde Q_L\cdot Q_L\Phi_L+
\tilde Q_R\cdot Q_R\Phi_R\right)}~,
\ee
for some constant $\lambda_{\phi}$,
where $\Phi_L$ is an $N\times N$ matrix chiral superfield that parametrizes the position
of the left semi-infinite $D4$-branes along the $NS$-brane (\ie\ in the $v$ direction).
This field is defined on the half-infinite line segment $x_6\le -L/2$, but what enters
$W_\phi$ is only its value at $x_6=-L/2$. Similarly, $\Phi_R$ parametrizes the position
in $w$ of the semi-infinite $D4$-branes on the right of figure \ref{origin}, $x_6\ge L/2$, and
what enters (\ref{extrasuppot}) is its value at $x_6=L/2$.

As mentioned above, the $U(N)\times U(N)$ global symmetry of the four dimensional theory
is part of the gauge symmetry of the five dimensional theory. Similarly, the vacuum expectation
values of the five
dimensional fields $\Phi_L$ and $\Phi_R$ correspond from the four dimensional point of view
to parameters in the Lagrangian -- turning them on gives masses to $(Q_L,\tilde Q_L)$ and $(Q_R, \tilde Q_R)$, respectively.

The 't Hooft coupling of the five dimensional theory,  $\lambda_N$ (see (\ref{lambdafour})),
has units of length. In section 2 we discussed the strong coupling regime, in which
this length is much larger than $l_s$. The gauge theory description is valid in the opposite
limit, $\lambda_N\ll l_s$. As usual, many aspects of the supersymmetric vacuum structure
are insensitive to the coupling, and can be compared between the two regimes.

Varying the superpotential $W=W_\lambda+W_\phi$ with respect to $\tilde Q_L$ gives the F-term equation
\be
\label{classfterm}\lambda Q_R\tilde Q_R\cdot Q_L+\lambda_\phi Q_L\Phi_L=0~.
\ee
Three similar equations are obtained by varying with respect to the other components of $Q$, $\tilde Q$.
Some of the solutions of (\ref{classfterm}) and of the D-term constraints can be described as follows.

\begin{figure}[t]
\begin{center}
\includegraphics[width= 100mm]{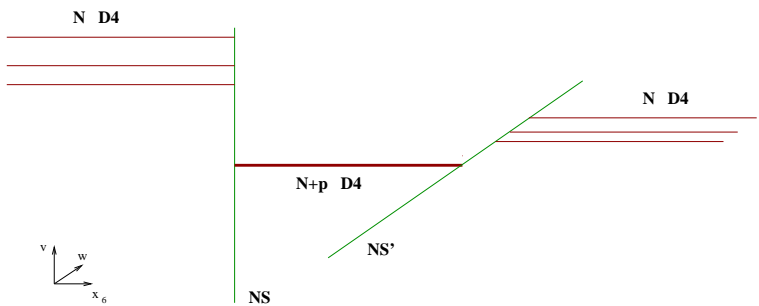}\end{center}
\caption{The brane realization of vacua with $\Phi_L,\Phi_R\not=0$.\label{vacuumone}}
\end{figure}

One branch of solutions is $Q_{L,R}=\tilde Q_{L,R}=0$, with $\Phi_L, \Phi_R$ general diagonal matrices.
The corresponding brane configuration is given in figure \ref{vacuumone}. As is clear both from the
field theory analysis and from the brane perspective, in this branch the chiral superfields are massive,
and  the theory generically reduces at low energies to $\cN=1$ pure SYM with gauge group $U(N+p)$.
This takes us back to the discussion of section 2.1.

Another branch of supersymmetric solutions corresponds to the brane configuration of figure
\ref{fullhiggs}. At strong coupling, the resulting vacua were discussed in section 2, around
equation (\ref{discU}).  At weak coupling, they can be described as follows.

In order for the $(N-p)$ $D4$-branes stretched from $-\infty$ to $\infty$ in figure \ref{fullhiggs} not
to intersect the $NS5$-branes, we must displace them in $(v,w)$ and/or $x_7$. In the gauge
theory,  the former corresponds to turning on non-zero expectation values for the five
dimensional fields $\Phi_L$, $\Phi_R$. Setting these expectation values to zero and displacing
the fourbranes in $x_7$ corresponds to turning on Fayet-Iliopoulos (FI) D-terms for the diagonal $U(1)$'s in
$U(N-p)$ subgroups of the two $U(N)$'s mentioned above. Denoting these D-terms by
$x_7^{L,R}$, the D-term potential sets
\bea
\label{FID}
Q^\dagger_L\cdot Q_L&=&x_7^L I_{N-p}~,\CR
Q^\dagger_R\cdot Q_R&=&x_7^R I_{N-p}~,
\eea
where we assumed that $x_7^{L,R}$ are positive, and set $\tilde Q_{L,R}=0$ for the $(N-p)$
flavors involved in (\ref{FID}). The latter is necessary for the F-term conditions (\ref{classfterm})
to be satisfied. Indeed, since we set $\Phi_L=0$ for the $(N-p)$ flavors in (\ref{FID}),  the F-term
constraint requires $\tilde Q_R\cdot Q_L$ to vanish. Since $Q_L$ is non-singular in a
$(N-p)\times (N-p)$ block constrained by (\ref{FID}), $\tilde Q_R$ must vanish in that block.
To solve (\ref{FID}), we take $Q_L=\sqrt{x_7^L}\delta^i_A$, where $i=1,\cdots, N-p$ is a flavor
index, and $A=1,\cdots, N-p$ a color one. Similarly, we take $Q_R=\sqrt{x_7^R}\delta^\alpha_A$,
where $\alpha$ is the flavor index, and the color index $A$ runs over the same range as for $Q_L$.
The D-term potential in the $U(N+p)$ four dimensional gauge theory, which requires
$Q^\dagger_LQ_L=Q^\dagger_RQ_R$, then implies that one must have $x_7^L=x_7^R$,
a constraint that is obvious from the geometry of figure \ref{fullhiggs}.

The above discussion takes care of $(N-p)$ colors and flavors. This leaves $2p$ colors
and $p$ flavors of $Q$, $\tilde Q$. To solve the D and F-term constraints, we take
$Q_L$, $\tilde Q_L$ to be diagonal and non-zero in a $p\times p$ block with
$i,A=N-p+1,\cdots, N$, and $Q_R$, $\tilde Q_R$ to be diagonal and non-zero in a
$p\times p$ block with  $\alpha,A=N+1,\cdots, N+p$. The F-term constraints (\ref{classfterm})
allow us to turn on an arbitrary $\Phi_L$ in the flavor sector with non-zero $Q_R$ and
vice-versa. Of course, the above forms of $Q$, $\tilde Q$ are up to gauge and global
transformations.

The above discussion involved vacua of the low energy gauge theory in which the
four dimensional $U(N+p)$ gauge group is completely broken. There are other branches
of the moduli space of supersymmetric vacua in which part of the gauge group is
unbroken. They can be described in the brane picture and in the low energy field theory
in a similar way.

In section  \ref{setups} we discussed a set of vacua which is described in the brane picture
by  the curved fivebrane (\ref{WittProf}), (\ref{NSUsol}) in the near-horizon geometry of  $N$
infinite $D4$-branes. We noted that the shape of the curved fivebrane involves in an important
way $g_s$ effects. Thus, the low energy field theory description should involve quantum
effects in the gauge theory. We next describe these vacua from the gauge theory point of view.

We start with the $U(N+p)$ gauge theory described in the beginning of this subsection.
To study its quantum dynamics, it is convenient to add to the theory two $N\times N$ massive
gauge singlet matrix chiral superfields $M_{LR}$, $M_{RL}$, and replace (\ref{suppot}) by
\bea
\label{bigsup}
W_\lambda=-{1\over\lambda}M_{LR}M_{RL}+M_{LR}\tilde Q_R\cdot Q_L
+M_{RL}\tilde Q_L\cdot Q_R~.
\eea
At low energies, we can integrate out the massive gauge singlets. Their equations
of motion set
\bea
\label{mesons}
M_{LR}=\lambda\tilde Q_L\cdot Q_R,\qquad M_{RL}=\lambda\tilde Q_R\cdot Q_L;
\eea
plugging this in (\ref{bigsup}) leads to (\ref{suppot}). Thus, the theory with superpotential
(\ref{bigsup}) is equivalent to the original one (\ref{suppot}) at low energies.

Consider now the theory (\ref{bigsup}).
We are interested in vacua in which the singlets $M_{LR}$, $M_{RL}$ are
non-singular matrices. Thus, the quarks $Q$, $\tilde Q$ are massive, and
can be integrated out. This gives an effective superpotential for
$M_{LR}$, $M_{RL}$, which can be calculated as follows. First, we use the
scale matching relation between the scale $\Lambda$ of the theory with
$2N$ flavors $Q$, $\tilde Q$ and the scale $\Lambda_L$ of the pure gauge
theory without them, which takes the following form in standard conventions:
\bea
\label{scalematch}
\Lambda^{N+3p}\det(M_{LR})\det(M_{RL})=\Lambda_L^{3(N+p)}~.
\eea
We then use the non-perturbative superpotential of the low energy pure
gauge theory,
\bea
\label{wwll}
W=(N+p)\Lambda_L^3~.
\eea
Combining this with the classical term in (\ref{bigsup}) we find the full
superpotential for $M_{LR}$, $M_{RL}$:
\bea
\label{quantsup}
W=-{1\over\lambda}M_{LR}M_{RL}+
(N+p)\left(\Lambda^{N+3p}\det(M_{LR})\det(M_{RL})\right)^{1\over N+p}~.
\eea
Varying with respect to (say) $M_{LR}$, and using the fact that the matrices $M_{LR}$,
$M_{RL}$ are non-degenerate, we find that
\bea
\label{formmlr}
M_{LR}M_{RL}=\lambda\left(\Lambda^{N+3p}\det(M_{LR})\det(M_{RL})\right)^{1\over N+p}~.
\eea
Solving this for the determinants, we find
\bea
\label{formdet}
\det(M_{LR})\det(M_{RL})=\lambda^{N(N+p)\over p}\Lambda^{N(N+3p)\over p}~.
\eea
In the vacua (\ref{formmlr}), the $SU(N)_L\times SU(N)_R$ global symmetry is
spontaneously broken to the diagonal $SU(N)$ by the expectation value of the
mesons $M_{LR}$, $M_{RL}$. This gives rise to a moduli
space of vacua labeled by the expectation value of the associated
Nambu-Goldstone bosons (which are linear combinations of the fields in $M_{LR}$
and $M_{RL}$).

We found that the quantum effects lead in this case to two (related) phenomena.
One is that the chiral superfields $Q_{L,R}$, $\tilde Q_{L,R}$ become massive;
their mass matrix,  $(M_{LR},M_{RL})$, is given by (\ref{formmlr}), (\ref{formdet}).
The other is the non-zero expectation value of $\tilde Q_L\cdot Q_R$, which breaks
the gauge symmetry $U(N+p)\to U(p)$. The scale of the breaking is proportional
to the gauge coupling and to the expectation value of $Q$, $\tilde Q$. Depending
on the parameters of the theory ($\Lambda$, $\lambda$, $g_{YM}$) one of the
effects occurs at a higher energy and dominates the dynamics.

If the scale of the breaking of the gauge symmetry is high, the low energy
theory can be thought of as  a pure $U(p)$ $\cN=1$ gauge theory with massive flavors.
The scale of this theory can be obtained by plugging (\ref{formmlr}) into
(\ref{quantsup}), which gives
\be
\label{nonescale}\Lambda_p^3=
\left(\Lambda^{N+3p}\lambda^N\right)^{1\over p}~.
\ee
In the brane system, the role of $\Lambda_p$ is played by $\xi/l_s^2$ (\ref{WittProf}).

The spectrum of the above gauge theory contains mesons whose masses are difficult to
compute, since this is non-holomorphic information. In section \ref{Tfotbaths} we will
see that in the regime where the brane description is valid, the masses of the flavor-singlet mesons can be
computed using the gravitational description.

\subsection{Compact $x_6$}

As mentioned above, another interesting brane configuration is obtained by compactifying
the $x_6$ direction on a circle, as in figure \ref{compactified}. We discuss here only
the supersymmetric case of periodic boundary conditions for the fermions on the circle. The
low-energy gauge theory in this case is an $SU(N+p)\times SU(N)$ four dimensional
${\cal N}=1$ supersymmetric gauge theory, with two chiral multiplets $A_i$ ($i=1,2$) in
the bi-fundamental representation and two ($B_i$, $i=1,2$) in the anti-bi-fundamental
\cite{Klebanov:1998hh,Klebanov:1999rd,Klebanov:2000nc,ks}. There is
also a superpotential generalizing (\ref{suppot}), of the form
\be
W = \lambda {\rm tr}(A_1 B_1 A_2 B_2 - A_1 B_2 A_2 B_1).
\ee

One can discuss this brane configuration in a decoupling limit which preserves the
five dimensional gauge dynamics including the Kaluza-Klein modes on the circle, or
take a different decoupling limit that keeps only the four dimensional gauge dynamics.
The latter limit was discussed in \cite{ks}, but we expect the moduli space to be
independent of precisely which limit we take.
As argued in  \cite{ks}  (see  \cite{Strassler:2005qs}
for a review), as one decreases the energy, this theory undergoes a series of ``duality
cascades'', such that the effective field theory describing physics at lower energy scales
is first a $SU(N)\times SU(N-p)$ theory, then a $SU(N-p)\times SU(N-2p)$ theory, and so on.
In the gravitational solution studied in \cite{ks}, this cascade eventually ends (if $N$ is a
multiple of $p$) in a confining background. This type of renormalization group flow implies
that also at high energies we cannot (once we take a decoupling limit from string theory)
discuss the theory at fixed $N$, since this theory has a Landau pole, but we have to
increase $N$ as we increase the UV cutoff,
and if we take the UV cutoff to infinity we must also take $N$ to infinity at the same time.

The moduli space of this theory was analyzed in detail in  \cite{Dymarsky:2005xt}, and we will just
quote their results here. They found that this theory actually has a large moduli
space of vacua. When $N$ is not a multiple of $p$ (and $p \ll N$), but rather $N \ {\rm mod}\ p = q$,
the theory has branches of the moduli space of (real) dimensions $6q$, $6(q+p)$, $6(q+2p)$, $\cdots$,
$6(q+np)$, $\cdots$,
which are each equivalent to a symmetric product of deformed conifolds (when the UV cutoff
goes to infinity, there is an infinite number of branches of this type). When $q=0$, the only
difference is that the dimension of the lowest branch is not zero but two, and on this lowest
branch some baryonic operators condense.

Seeing this moduli space in the Klebanov-Strassler description, corresponding to the decoupling
limit of the four dimensional gauge theory when this theory is strongly coupled, is straightforward but not
completely trivial; the different branches involve $(q+np)$ $D3$-branes that are free to move on the
conifold and that generally back-react on its geometry. We can see precisely the same moduli
space in the gravitational description of the previous section, which is valid in the opposite
limit in which the five dimensional gauge dynamics is strongly coupled. We saw in the previous
section that in this description the background involves a spiralling fivebrane, which reduces the
rank of the gauge group as we go down in the radial direction, just as in the Klebanov-Strassler
cascade (but here this reduction comes from explicit brane sources, rather than from background
fluxes). We also saw that for given UV boundary conditions we have an infinite set
of vacua (\ref{xinn}), which differ in the value of the extreme IR flux (we identify the vacuum in
which the IR flux is equal to $(q+np)$ with the $6(q+np)$-dimensional branch of the moduli space
mentioned above). Note that, unlike in the case of Klebanov and Strassler, in our case we do not
have a good description of the vacua with the lowest values of $n$, where back-reaction of the
five-brane becomes important in our solutions. However, we do have good descriptions for all the
vacua with $p \ll q+np$.

\section{ Observables and spectra}\label{Tfotbaths}

In the decoupling limit described in section \ref{holemb}, holography
relates non-normalizable modes in the bulk to operators in the dual
field theory, and the bulk path integral with sources to the generating
functional of correlation functions of the corresponding field theory operators.
Normalizable states in the bulk correspond to states in the dual field theory.

In this section we discuss some aspects of this map for operators and states
associated with the brane intersection, which correspond to fields living on
the curved $NS5$-brane.\footnote{There are also ``bulk'' modes
associated with the field theory living on the infinite $D4$-branes. In
the probe approximation (\ie\ to leading order in $p/N$) they
are not affected by the extra fivebrane.} We start by identifying modes of the
fivebrane with operators in the low energy field theory, and then move on to calculate
the spectrum of mesons. In section \ref{TsotNGm} we discuss the potential existence
of normalizable Nambu-Goldstone modes associated with global symmetry breaking.

\subsection{Operator matching}

The curved $NS5$-brane (\ref{WittProf}) has a two component boundary, corresponding
to large positive $x_6$ (and hence large $w$ (\ref{asshape})), and large negative $x_6$
(and large $v$). The latter corresponds to going to the boundary at large $u$ along
the $NS$-brane; the former, along the $NS'$-brane. We will focus, for concreteness,
on the operator map for modes defined  on the $NS$-brane. The
discussion of modes living on the $NS'$-brane is very similar.

To get a qualitative guide to the spectrum of operators on the $NS$-brane, it is
convenient to go back to figure \ref{origin}, which does not take into account
$g_s$ effects, but is nevertheless useful. In this limit, the
$NS$-brane fills the $v$ plane; the bosonic modes living on it are scalar fields
describing its fluctuations in the transverse directions\footnote{From the ten
dimensional point of view, one of these scalar fields, ${\rm Im}(z)=x_{11}$, is
non-geometric.} $(w,z,x_7)$, and a self-dual 2-form field. Translations of the fivebrane
in $z$ change the coupling of the $U(N+p)$ gauge theory. Hence, the bulk
operator corresponding to the relative location of the $NS$ and $NS'$-branes
couples to the gauge theory Lagrangian. We are mainly interested in mesons,
and will not discuss this mode in detail.

The scalar fields $w$ and $x_7$ couple to gauge-invariant operators constructed out
of the chiral superfields $Q_L$, $\tilde Q_L$ living at the intersection of the
$D4$-branes and the $NS$-brane. To study these operators in the bulk, we write
\bea
x_7=\epsilon{\tilde x}_7(v,\vb,x^\mu),\quad
w=w_0(v)+ \epsilon{\tilde w}(v,\vb,x^\mu),
\eea
where $w_0$ is the undeformed profile (\ref{WittProf}), and we work to leading
order in the deformations ${\tilde x}_7$, ${\tilde w}$. Specializing to
wavefunctions with well defined charge (\ref{branesym}),
\bea\label{PertIdent}
{\tilde w}=v^n{\tilde w}^{(n)}(v{\bar v}, x^\mu),\quad {\tilde x}_7=v^n{\tilde x}_7^{(n)}(v{\bar v}, x^\mu),
\eea
and plugging into the fivebrane action, we find the linearized
equation of motion
\bea\label{UnivKG}
\left(\d_{x_6}^2+\partial_\mu\partial^\mu\right) F-\frac{n^2}{\lambda_p^2}F=0,
\eea
where $F={\tilde w}^{(n)}, {\tilde x}_7^{(n)}$, and we used $x_6$ rather than $v$ to parameterize the
worldvolume of the unperturbed fivebrane, (\ref{WittProf}).

The field theory operators corresponding to (\ref{PertIdent}) can be identified
by matching the transformation properties under the symmetries\footnote{Of course, all the states
and operators we discuss are singlets of the $SU(N)_L\times SU(N)_R$ symmetry, since this is
a gauge symmetry in our setup.}. This leads to
\bea\label{opmatch}
&&v^n{\tilde w}^{(n)}\leftrightarrow
Q_L \Phi_L^n {\tilde Q}_L,\nonumber\\
&&v^n{\tilde x}_7^{(n)}\leftrightarrow
Q_L \Phi_L^n {Q}^\dagger_L-{\tilde Q}^\dagger_L \Phi_L^n {\tilde Q}_L.
\eea
So far, we have focused on operators associated with the boundary along the
$NS$-brane, at large negative $x_6$. Of course, perturbations introduced
there propagate to large positive $x_6$ (where the equations of motion (\ref{UnivKG})
are corrected in a way that will be described below). In order to study
perturbations containing only the operators (\ref{opmatch}), and not their
analogs with $L\to R$, one needs to choose solutions of the equations of motion
that decay rapidly as $x_6\to\infty$.

At first sight, for $n>0$ the operators on the right-hand side of (\ref{opmatch}) seem to
involve the five dimensional modes living on the infinite $D4$-branes. However, one
can use  the equations of motion of $Q$, $\tilde Q$  (\ref{classfterm})  to
express them purely in terms of fields in the four dimensional low energy theory.
For example, the  ${\tilde w}^{(1)}$ mode corresponds to the operator
$\tilde Q_L\cdot Q_R\tilde Q_R\cdot Q_L$. For higher $n$, one finds operators of
the schematic form $\tilde Q_L Q_L (\tilde Q_R Q_R)^n$. A similar set of operators
with $L\leftrightarrow R$ is obtained from the other boundary, $x_6\to\infty$.

To analyze the solutions of (\ref{UnivKG}), it is convenient to write the operators
(\ref{opmatch}) in momentum space, \ie\ take $F(x_6, x_\mu)=F(x_6)\exp(ik^\mu x_\mu)$.
For $n>0$ and sufficiently small $k^2$, the solutions of (\ref{UnivKG}) are in general
non-normalizable as $x_6\to-\infty$; this gives rise to the sources holographically
related to the operators in the low energy theory at the intersection, (\ref{opmatch}).
For $n=0$ and timelike (or null) momentum, $k^2\le 0$,  the solutions of
(\ref{opmatch}) are in fact delta-function normalizable; they correspond to $4+1$
dimensional scattering states which are not localized at the intersection. The same is true
for non-zero $n$ and  $-k^2 \geq n^2 / \lambda_p^2$. Thus, we conclude that the map
(\ref{opmatch}) is only valid for $n\ge 1$, and that in addition to the field theory
modes, the
brane system contains a continuum of four dimensional states above the gap $n/\lambda_p$.
For $n=0$, this continuum starts at zero energy.\footnote{The fact that the map (\ref{opmatch})
is not valid for $n=0$ is natural from the field theory perspective, since the operators
on the right-hand side of (\ref{opmatch}) are analogs of the $U(1)$ part of $U(N)$ in
other holographic systems.}

Having a continuum with some discrete localized states is natural in field theories
with defects. In our case we can think of the brane intersection region as a codimension
one defect inside the fivebrane; the discrete states are localized near the intersection,
while the continuum corresponds to states propagating in the bulk of the fivebrane. In
the limit we take, the four dimensional fields are not decoupled from the higher
dimensional ones.

The continuous spectrum in our system is also somewhat similar to what happens in Little
String Theory, where
in a given charge sector one typically finds a discrete spectrum of localized modes and
a continuous spectrum corresponding to modes propagating in an asymptotically linear
dilaton throat (see \eg\ \cite{Aharony:2004xn}). However, in that case the continuum does
not have a simple interpretation in terms of a local field theory on the fivebranes.

In the next subsection we turn to the spectrum of normalizable modes of the confining
vacuum (\ref{WittProf}) of holographic MQCD. These modes correspond to particles in
the dual gauge theory. We will focus on the scalars corresponding to transverse
fluctuations of the fivebrane. Parametrizing the worldvolume as in section \ref{setups},
we can label these directions by $v, w$ and $x_7$. Since the classical shape (\ref{WittProf})
is localized at $x_7=0$, it is easiest to study the fluctuations in the $x_7$ direction, which
are decoupled from those in the other directions. Thus, we start with those, and then move
on to the other ones.

We find a discrete spectrum of massive states, below the continuum discussed
above. If we take a limit that decouples the higher dimensional fields from the four dimensional
field theory, the continuous part of the spectrum should decouple, leaving behind a set of
discrete states. We assume that the discrete states in this limit are continuously related to the
ones we find here, but it is also possible that as we interpolate between the two regimes, some states
disappear into the continuum while other states emerge from it.

\subsection{ Spectrum of $x_7$ fluctuations}

We start with the analysis of the fluctuations along the $x_7$ direction
 which is transverse to the
$D4$, $NS$ and $NS'$-branes. As mentioned above, these  fluctuations  are decoupled from
those corresponding to the other directions, and therefore their analysis  is more tractable. The
corresponding gauge theory meson operators are described in the previous subsection (see
(\ref{opmatch}) and the discussion around it).

Before incorporating the fluctuations, the $x_7$ position of the fivebrane is a constant, $x_7= X_7$
(which is one of the components of $\vec{r_0}$ in (\ref{TheMetric})).
Perturbing around the classical solution, we have
\be\label{pertxseven}
x_7= X_7 + \epsilon  \tilde x_7(z,\bar z,x^\mu).
\ee
Consider first the case  $X_7=0$. Plugging (\ref{pertxseven}) into the background  (\ref{TheMetric})
\bea
ds^2&=&H^{-1/3}[dx_{\mu}^2+dzd{\bar z}]+
H^{2/3}[dvd{\bar v}+dwd{\bar w}+dx_7^2],\nonumber\\
C_6&=&\frac{1}{2i}H^{-1}d^4x\wedge dz\wedge d{\bar z},\qquad
H=\frac{\pi \lambda_N l_s^2}{(v{\bar v}+w{\bar w}+x_7^2)^{3/2}}\,,
\eea
we find the induced metric on the probe branes,
\bea\label{gabinduced}
ds^2=H^{-1/3}\left(dx_i^2+g_{ab}dx^adx^b\right),\qquad x^a=(z,\zb,t).
\eea
In (\ref{gabinduced}) and below we take (without loss of generality) the field $\tilde x_7$
to depend only on the time $t$ and not on the spatial coordinates in $\IR^{3,1}$.
Keeping only  terms  up to quadratic order in $\eps$, we find (denoting $\pa \equiv \pa / \pa z$,
$\db = \pa / \pa {\bar z}$)
\bea
g=\left(\begin{array}{ccc}
  \eps^2 H \pa \tilde x_7 \pa \tilde x_7  & g_{z{\bar z}} & \eps^2 H \pa \tilde x_7\dot {\tilde x}_7 \\
 g_{z{\bar z}} &  \eps^2 H \db \tilde x_7 \db \tilde x_7  & \eps^2 H \db \tilde x_7\dot {\tilde x}_7
\\
\eps^2 H \pa \tilde x_7\dot {\tilde x}_7 & \eps^2 H \db \tilde x_7\dot {\tilde x}_7&
-1+H\eps^2(\dot {\tilde x}_7)^2
\end{array}\right).
\eea
Here we defined
\bea
g_{z{\bar z}}=\frac12[1+  H (|\pa v|^2+ |\pa w|^2)] +\eps^2 H |\pa \tilde x_7|^2.\nonumber
\eea
The harmonic function takes the form
\be
H=\frac{\pi \lambda_N l_s^2}{(u_{cl}^2+ x_7^2)^{3/2}}=
\frac{\pi \lambda_N l_s^2}{\left[2\xi^2\cosh\left(\frac{2x_6}{\lambda_p}\right)+  x_7^2\right]^{3/2}}\,.
\ee
We also define the dimensionless coordinates $\hat x_7=\frac{ \tilde x_7}{\sqrt{2}\xi}$,
$\hat x_6=\frac{x_6}{\lambda_p},\ \hat x_{11} =\frac{x_{11}}{\lambda_p},\ \hat t =\frac{t}{\lambda_p}$,
and the dimensionless ratio $q$
\be\label{defq}
q\equiv \frac{\sqrt{2}\xi\lambda_p^2}{\pi \lambda_N l_s^2}.
\ee
Note that the condition (\ref{xinearh}) implies that $q\gg (\lambda_p/\lambda_N)^2=(p/N)^2$. However,
this lower bound on $q$ is very weak in the probe limit $p\ll N$. E.g., if we keep the ``QCD scale''
$\xi/l_s^2$ fixed in units of the five dimensional gauge coupling $\lambda_p$ or $\lambda_N$, we
have $q \ll 1$ in the probe limit.

In terms of the above definitions and omitting the hats,
expanding to second order in $\epsilon$ we find in this case the following Lagrangian
(up to a multiplicative constant)
\bea\label{Lagx7}
L&\simeq&\int d^2z H^{-1}(\sqrt{\frac{1}{2}\mbox{det}(2g)}-1)\simeq \CR
&\simeq&  \epsilon^2\int d^2z \left[ 2 \pa x_7 \db x_7 - \frac{1}{2}\left ( 1+ \frac{1}{q \sqrt{\cosh(2x_6)}}\right ) \dot x_7^2 \right]. \CR
\eea
The corresponding equation of motion for $x_7(x_6,x_{11},t)$ is
\be\label{mx7}
-\pa^2_{x_6} x_7- \pa^2_{x_{11}}x_7 +  \left ( 1+ \frac{1}{q \sqrt{\cosh(2x_6)}}\right ) \ddot x_7=0.
\ee
If we look at a mode of fixed mass and momentum in $x_{11}$, such that
$ x_7(x_6,x_{11},t)=e^{i n x_{11} + i m t} x_7(x_6)$, we get the following
equation\footnote{Note that the charge of the state under the global $U(1)$
symmetry (\ref{branesym}) is $-n$.}
\be\label{eomx7}
\pa^2_{x_6} x_7 - \left[ n^2- m^2 \left ( 1+ \frac{1}{q \sqrt{\cosh(2x_6)}}\right )\right ]x_7 =0.
\ee

In the asymptotic region of large $|x_6|$, the equation we find is (\ref{UnivKG})
\be
\pa^2_{x_6} x_7 = (n^2 - m^2) x_7.
\ee
Clearly, for $m^2 > n^2$ the solutions in this region are just plane waves, while for $m^2 < n^2$
the solutions decay or grow exponentially at infinity. We can think of (\ref{eomx7}) as a
Schr\"odinger equation describing the motion of a particle with vanishing energy in the potential
\be\label{effpot}
V(x_6)=n^2- m^2 \left ( 1+ \frac{1}{q \sqrt{\cosh(2x_6)}}\right ).
\ee
This makes it clear that when $m^2 > n^2$ the solutions describe scattering off a potential well
and there is a delta-function normalizable solution for any value of $m^2$, while for $m^2 < n^2$
the solutions describe bound states in a potential so we expect to get solutions only for discrete
values of $m^2$ (one can show that at least one bound state exists for all $q$ and $n$).

Going back to dimensionful variables, we see that we have a continuum of states with
$m > n / \lambda_p$ , and a discrete spectrum for $m < n / \lambda_p$. As discussed
above, the discrete spectrum can be thought of as describing mesons in the confining
vacuum of the gauge theory of section 3 at strong coupling, while the continuum is
associated with higher dimensional modes.

\begin{figure}[t]
\begin{center}
\includegraphics[width= 90mm]{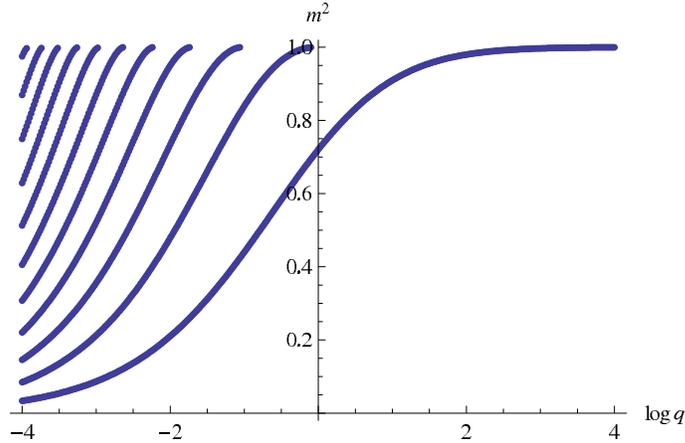}\end{center}
\caption{ The spectrum of $x_7$ fluctuations: $m^2$ (in units of $1/\lambda_p^2$) as a function of $q$.\label{spectrumx7}}
\end{figure}

To examine the spectrum of mesons, we numerically solved (\ref{mx7}) for the case $n=1$.
The resulting spectrum is depicted in figure \ref{spectrumx7} as a function of $q$. The
qualitative features of this spectrum are obvious by thinking about the analogous
Schr\"odinger problem. For small $q$, the potential well (\ref{effpot}) is very deep,
so we expect many bound states, with the low lying ones having $m^2\propto q/\lambda_p^2$. On the other
hand, for large $q$ the potential is very shallow, so we expect (and find) just a single
bound state, very close to $m=1/\lambda_p$.

Note that the potential (\ref{effpot}) is symmetric under $x_6\to -x_6$, which acts as charge
conjugation on the vector superfields, and exchanges $Q_L\leftrightarrow Q_R$. The minimum
of the potential is at $x_6=0$, and it monotonically increases with $|x_6|$. The mesons depicted in
figure \ref{spectrumx7} are localized near $x_6=0$; they can be thought of as having significant
overlap with both $Q_L$ and $Q_R$.

For $n>1$ the states described by (\ref{eomx7}) can be thought of as ``exotics'' since
they have the $U(1)$ charge of $Q_L^\dagger Q_L (\tilde Q_R Q_R)^n$. Their spectrum is
similar to that of figure \ref{spectrumx7}, but the masses are larger. This can be
seen by noting that to have zero energy bound states, the potential (\ref{effpot}) has
to be negative at the origin. This implies that the bound state masses $m$ always satisfy the bound
\be\label{boundmass}
(m\lambda_p)^2>{n^2\over 1+{1\over q}}~.
\ee
Thus, the masses of ``exotics'' containing $n$ pairs of $\tilde Q_R Q_R$ and/or
$\tilde Q_L^\dagger Q_L^\dagger$ grow with $n$.

For the ``massive'' case, where the $N$ $D4$-branes are displaced from the $p$ $D4$-branes
that stretch between the $NS$ and $NS'$-branes  in the $x_7$ direction by the distance $X_7$
(see figure \ref{massive}), the harmonic function takes the form
\be
H= \frac{\pi \lambda_N l_s^2}{\left[2\xi^2\cosh\left(\frac{2x_6}{\lambda_p}\right)+
(X_7+\eps \tilde x_7)^2\right]^{3/2}}~.
\ee
The equation of motion reads (after rescaling as above, with $X_7$ rescaled in the same way as
$\tilde x_7$)
\be\label{mx7b}
\pa^2_{x_6} x_7 - \left[ n^2- m^2 \left ( 1+ \frac{\cosh(2x_6)}{q(\cosh(2x_6) + X_7^2)^{3/2}}\right )\right ]x_7 =0.
\ee
The potential (\ref{effpot}) takes in this case the form
\be\label{neweffpot}
V(x^6)=n^2- m^2 \left ( 1+ \frac{\cosh(2x_6)}{q(\cosh(2x_6) + X_7^2)^{3/2}}\right ).
\ee
For $X_7<1/\sqrt2$ or, in terms of dimensionful variables, $X_7<\xi$,  the potential
(\ref{neweffpot}) is qualitatively similar to (\ref{effpot}) -- it has a unique minimum at
$x_6=0$ and no other extrema. For $X_7>1/\sqrt2$, the extremum at $x_6=0$ becomes
a local maximum, and the potential  (\ref{neweffpot}) becomes a double well potential,
with minima at the two solutions of the equation $\cosh(2x_6)=2X_7^2$. As $X_7$
increases, the potential becomes more and more sharply peaked (for sufficiently
small $q$), and the two minima move to large $|x_6|$.

For large $X_7$ and small $q$, the solutions of the Schr\"odinger problem that gives the
masses, (\ref{mx7b}), split into wavefunctions localized in the two wells. If this was the
end of the story, the theory would break the charge conjugation symmetry $x_6\to -x_6$,
and the mass spectrum would split into degenerate doublets related by the symmetry.

\begin{figure}[!]
\begin{center}
\includegraphics[width= 80mm]{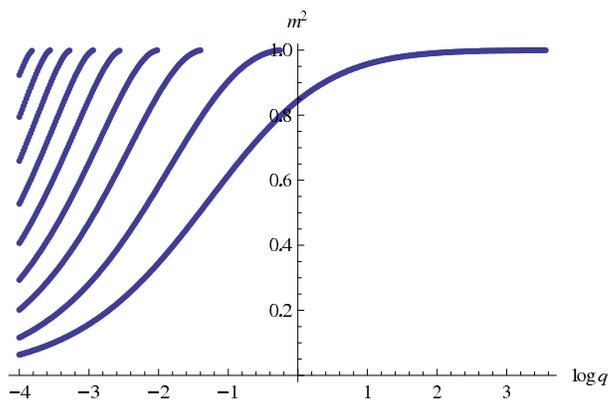}\end{center}
\caption{ The spectrum of $x_7$ fluctuations for $X_7^2=1$.\label{spectrum1}}
\end{figure}

\begin{figure}[!]
\begin{center}
\includegraphics[width= 80mm]{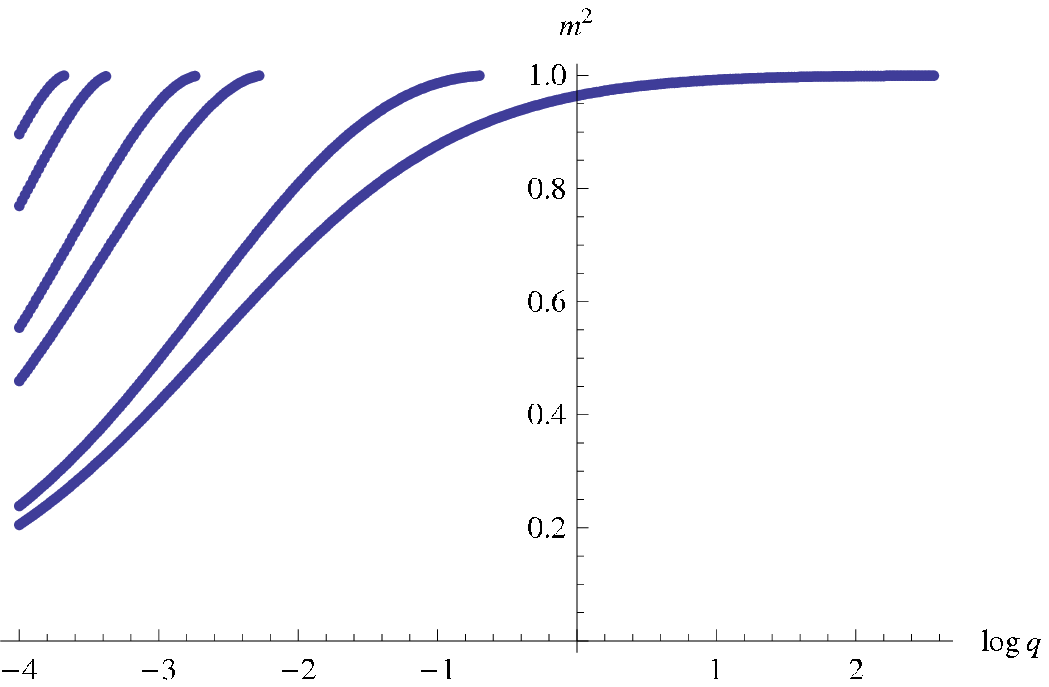}\end{center}
\caption{ The spectrum of $x_7$ fluctuations for $X_7^2=10$.\label{spectrum10}}
\end{figure}

\begin{figure}[!]
\begin{center}
\includegraphics[width= 80mm]{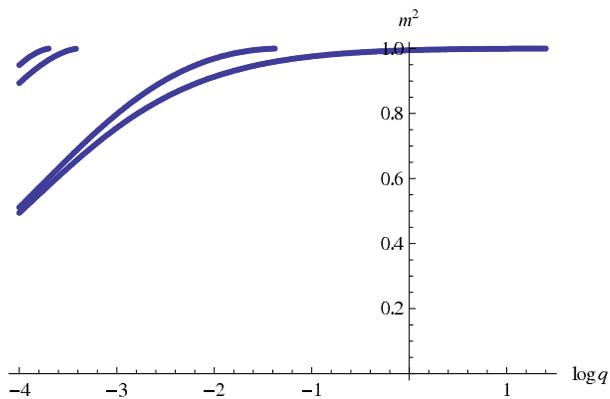}\end{center}
\caption{ The spectrum of $x_7$ fluctuations for $X_7^2=100$.\label{spectrum100}}
\end{figure}

Of course, as is well known, symmetry breaking does not happen in quantum mechanics due
to tunneling. Rather than exactly degenerate doublets, we expect to find approximately degenerate
pairs of states corresponding to the sum and difference of wavefunctions localized in the
left and right wells. In figures \ref{spectrum1}-\ref{spectrum100} we present numerical results
for the mass spectrum for three values of $X_7$ in the ``broken phase'', $X_7^2=1,10,100$.
As one can see from these figures, for fixed $X_7$ the splitting of the spectrum into
approximately degenerate pairs becomes more and more pronounced as $q$ decreases,
while for fixed $q$ it becomes more pronounced as $X_7$ increases, as one would expect.

From the point of view of the low energy field theory of section 3, the change in
the nature of the wavefunctions of the normalizable states described above means
that as $X_7$ increases, the mesons go from having significant overlaps with
both $Q_L$ and $Q_R$ to being primarily made out of $Q_L$ (for wavefunctions
localized in the left well) and of $Q_R$ (for those in the right well). These
``flavor eigenstates'' have small mixings, such that the mass eigenstates are
the sum and difference of the mesons made out of $Q_L$ and those made out of $Q_R$.

\subsection{Fluctuations of the other transverse directions}\label{holfluc}

Next, we analyze the transverse fluctuations in some of the other directions.
From the field theory point of view, one expects to find a qualitatively
similar spectrum to that described in the previous subsection. As we will
see, in this case the equations for the fluctuations are coupled, and thus
are more difficult to solve numerically. However, in certain limits we will be
able to solve them, and find that the expectations are realized.

As before, to study the small fluctuations we expand around the holomorphic profile,
\bea
w=w_0(z)+\eps w_1(z,{\bar z},t),\qquad v=v_0(z)+\eps v_1(z,{\bar z},t).
\eea
Similar to  (\ref{gabinduced})  we determine the  nontrivial part  of $g_{ab}$ in the induced metric.
To compute the determinant of $g_{ab}$ to quadratic order in $\eps$, we should keep quadratic terms
in $g_{z{\bar z}}$ and $g_{tt}$, and linear contributions to the other components of $g_{ab}$:
\bea
g=\left(\begin{array}{ccc}
\eps H(\d w_0\d \wb_1+\d v_0\d \vb_1)&g_{z\bar z}&
\frac{H\eps}{2}(\d w_0{\dot {\bar w}_1}+\d v_0{\dot \vb_1})\\
g_{z\bar z}&\eps H(\db \wb_0\db w_1+\db \vb_0\db v_1)&
\frac{H\eps}{2}(\db {\bar w}_0 {\dot w}_1+\db {\bar v}_0{\dot v_1})\\
\frac{H\eps}{2}(\d w_0{\dot \wb_1}+\d v_0{\dot \vb_1})&\frac{H\eps}{2}(
\db {\bar w}_0 {\dot w}_1+\db {\bar v}_0{\dot v_1})&
H\eps^2({\dot w}_1{\dot\wb}_1+{\dot v}_1{\dot\vb}_1)-1
\end{array}\right).
\nonumber
\eea
Here
\be
2g_{z\bar z}=1+H[\d (w_0+\eps w_1)\db (\wb_0+\eps\wb_1)+\d (v_0+\eps v_1)\db (\vb_0+\eps\vb_1) +
\eps^2 \db w_1 \d \wb_1 + \eps^2 \db v_1 \d \vb_1].
\ee
Next we compute
\bea
\mbox{det}(2g)&=&2\left[(2g_{z\zb})^2(-g_{tt})-4(\eps H)^2(\db \wb_0\db w_1+\db \vb_0\db v_1)
(\d w_0\d \wb_1+\d v_0\d \vb_1)\right]
\nonumber\\
&&+2(\eps H)^2(\d w_0{\dot \wb_1}+\d v_0{\dot \vb_1})(
\db {\bar w}_0 {\dot w}_1+\db {\bar v}_0{\dot v_1})(2g_{z\zb}),
\eea
so that
\bea
&&\sqrt{\frac{1}{2}\mbox{det}(2g)}=(2g_{z\zb})[1-\frac{H\eps^2}{2}({\dot w}_1{\dot\wb}_1+{\dot v}_1{\dot\vb}_1)]\nonumber\\
&&\qquad+\frac{2(\eps H)^2}{1+H(\d w\db \wb+\d v\db \vb)}\left[-
(\db \wb_0\db w_1+\db \vb_0\db v_1)
(\d w_0\d \wb_1+\d v_0\d \vb_1)\right]\nonumber\\
&&\qquad+\frac{(\eps H)^2}{2}(\d w_0{\dot \wb_1}+\d v_0{\dot \vb_1})(
\db {\bar w}_0 {\dot w}_1+\db {\bar v}_0{\dot v_1}).
\eea
The Lagrangian density in $\IR^{3,1}$ becomes
\bea
L&=&\int d^2z \frac{1}{H}(\sqrt{\frac{1}{2}\mbox{det}(2g)}-1)\nonumber\\
&=&
\int d^2z\left[\d (w_0+\eps w_1)\db (\wb_0+\eps\wb_1)+\d (v_0+\eps v_1)\db (\vb_0+\eps\vb_1)
-\eps^2 g_{z\zb}({\dot w}_1{\dot\wb}_1+{\dot v}_1{\dot\vb}_1)\right.
\nonumber\\
&&\qquad\qquad-
\frac{2\eps^2 H}{1+H(\d w\db \wb+\d v\db \vb)}
(\db \wb_0\db w_1+\db \vb_0\db v_1)
(\d w_0\d \wb_1+\d v_0\d \vb_1)\nonumber\\
&&\qquad\qquad\left.+\frac{\eps^2 H}{2}(\d w_0{\dot \wb_1}+\d v_0{\dot \vb_1})(
\db {\bar w}_0 {\dot w}_1+\db {\bar v}_0{\dot v_1}) +
\eps^2 \db w_1 \d \wb_1 + \eps^2 \db v_1 \d \vb_1
\right].
\eea
The terms linear in $\eps$ do not contribute to the equations of motion for $w_1$ and $v_1$, and since $H$ appears only
multiplying $\epsilon^2$, we can replace it by its value $H_0$ in the original solution. Dropping an overall power of $\eps$, we find
\bea\label{QuadrLagr}
L&=&\int d^2z\left[\d w_1\db \wb_1+\d v_1\db \vb_1
+ \db w_1 \d \wb_1 + \db v_1 \d \vb_1 -\right.\nonumber\\
&& \qquad\qquad \left. \frac{2H_0}{F}(\db \wb_0\db w_1+\db \vb_0\db v_1)
(\d w_0\d \wb_1+\d v_0\d \vb_1)\right]\\
&& \qquad\qquad -\frac{1}{2}\int d^2z \left[F
(|\dot w_1|^2+|\dot v_1|^2)-H_0(\d w_0{\dot \wb_1}+\d v_0{\dot \vb_1})(
\db {\bar w}_0 {\dot w}_1+\db {\bar v}_0{\dot v_1})\right]\nonumber,
\eea
where we defined
\bea
F\equiv 1+H_0(\d w_0\db \wb_0+\d v_0\db \vb_0).
\eea
For the profile (\ref{WittProf}) we have
\bea
v_0=\xi e^{-z/\lambda_p},\quad w_0=\xi e^{z/\lambda_p},\quad
H_0=\frac{\pi \lambda_Nl_s^2}{(v_0\vb_0+w_0\wb_0)^{3/2}},\quad
F=1+\frac{2\xi^2}{\lambda_p^2}H_0\cosh\left(\frac{z+\zb}{\lambda_p}\right).
\eea
The equations of motion that follow from (\ref{QuadrLagr}) are complicated. To get
some insight about the structure of the spectrum we first analyze the asymptotic
form of the equations of motion at large values of $|{\rm Re}(z)|= |x_6|$.
Keeping only the leading order terms, we arrive at an approximate asymptotic Lagrangian:
\bea\label{SimpleAct1}
L\approx\int d^2z\left[\d w_1\db \wb_1+\d v_1\db \vb_1+ \db w_1 \d \wb_1 + \db v_1 \d \vb_1
\right]
-\frac{1}{2}\int d^2z
\left[|\pa_\mu w_1|^2+|\pa_\mu v_1|^2\right],
\eea
where for completeness we have restored the full $x_\mu$ dependence.
The corresponding equations of motion read
\bea\label{KleinG1}
\d\db w_1-\frac{1}{4}{\pa^\mu\pa_\mu w}_1=0,\qquad \d\db v_1-\frac{1}{4}{\pa^\mu\pa_\mu v}_1=0.
\eea
Taking a fixed momentum $n$ in $x_{11}$ and a fixed four dimensional mass $m$ as above, the resulting equation of motion is
\bea
\d_{x_6}^2 w_1-\left(\frac{n^2}{\lambda_p^2}-m^2 \right ) w_1=0,
\eea
and similarly for the other modes. Thus, as in the previous section, we have
a discrete spectrum  for  $m<|n|/\lambda_p$, and a continuum for $m>|n|/\lambda_p$.
Note that since in this case $v$ and $w$ themselves carry a $U(1)$ charge $\pm 1$,
these fluctuations carry $U(1)$ charges $n \pm 1$.

Let us describe in more detail the modes coming from fluctuations of the absolute values of
$v$ and $w$, namely
\be
V= |v|= V_0+
\epsilon V_1, \qquad W= |w|= W_0+ \epsilon W_1.
\ee
In terms of the coordinates $u$ and $\alpha$ defined in (\ref{SpecEqn})
\bea\label{UWvar}
ue^{i\alpha}=(V_0+\epsilon V_1)+i(W_0+\epsilon W_1).
\eea
Substituting these variables into the Lagrangian  (\ref{QuadrLagr}) we get
\bea\label{LagrMs10}
L&=&\frac{1}{2}\int dx_6\left[(V_1')^2+(W_1')^2+(V_1^2+W_1^2)-
{H_0}\frac{[V_0'V_1'+W_0'W_1'-V_0V_1-W_0W_1]^2}{(\lambda_p^2+u^2 H)_0}\right]\nonumber\\
&&+\int dx_6
\frac{1}{2}(\lambda_p^2+u^2 H)_0[(\d_\mu V_1)^2+(\d_\mu W_1)^2].
\eea
Here we have (after rescaling $x_6$ as before)
\bea
V_0=\xi e^{-x_6},\quad W_0=\xi e^{x_6},\quad (\lambda_p^2+u^2 H)_0=\lambda_p^2+\frac{Q}{u}=
\lambda_p^2+\frac{Q}{\xi(e^{2x_6}+e^{-2x_6})^{1/2}}.\nonumber
\eea
Assuming that $W_1\sim \sin(mt)$, $V_1\sim \sin(mt)$, we find the Lagrangian
\bea\label{LagrMs11a}
L&=&\frac{1}{2}\int dx_6\left[(V_1')^2+(W_1')^2+(V_1^2+W_1^2)-[\lambda_p^2+u^2H]_0[(m V_1)^2+(m W_1)^2]
\right]
\nonumber\\
&&-\frac{1}{2}\int dx_6\frac{H_0}{(\lambda_p^2+u^2H)_0}[W_0W_1'-V_0V_1'-V_0V_1-W_0W_1]^2.
\eea
This Lagrangian is still too complicated to analyze. In the Appendix we invoke several
approximations that enable us to simplify its form for some range of parameters.
Using these approximations we get a decoupled  Lagrangian  for
$V_+\equiv \frac{1}{\sqrt{2}}( V_1+W_1)$ of the form
\bea\label{WKBlagr}
L&\approx&\frac{1}{2}\int dx_6 \left[(V_+')^2+
\left(1-(m \lambda_p)^2-\frac{(m \lambda_p)^2}{q}\frac{1}{\sqrt{\cosh(2x_6)}}\right)V_+^2\right].
\eea
This is exactly the same Lagrangian as the one we found (with no assumptions and approximations)
for the fluctuations along $x_7$ (\ref{Lagx7}). The solution of the corresponding equation of motion is
thus identical  (for $n=1$) to the  numerical solution described in figure \ref{spectrumx7}.

\subsection{Self-dual $B$ field}\label{TsotNGm}

The remaining bosonic field living on the curved fivebrane is the self-dual 2-form $B$ field.
Its fluctuations give rise to a spectrum of mesons, whose masses can be analyzed
as in the previous subsections.  We will not discuss the details of this analysis here.

Instead, we will comment briefly on the following question. In the original brane
configuration of figure \ref{origin}, there are in fact two independent self-dual $B$ fields,
living on the two $NS5$-branes.  One can think of them as generating two $U(1)$
symmetries in the low energy field theory. The corresponding gauge fields
are obtained by reducing the self-dual $B$-field living on the $NS$-brane
on the angular direction in the $v$-plane, and similarly for the $NS'$-brane.

In the confining vacuum described by (\ref{WittProf}), the two $NS5$-branes
connect, and these symmetries are broken to the diagonal subgroup,
$U(1)_L\times U(1)_R\rt U(1)_D$. Superficially, the situation seems to be very
similar to that in the Sakai-Sugimoto model \cite{Sakai:2004cn} (a closer
analog to our situation is the non-compact analog of that model, studied in
\cite{Antonyan:2006vw}). In that case, the symmetry breaking was spontaneous,
and gave rise to a massless pion, which corresponded to the zero mode of the
component of the gauge field on the flavor $D8$-branes along the $U$-shaped
brane. In analogy, one might expect that in our system a massless pion would
arise from the self-dual $B$ field with components along the curved fivebrane.
However, there are two important differences between our case and the
Sakai-Sugimoto model. In our case the brane only approaches the boundary at
infinite values of $x_6$; also, in our case the radius of the $x_{11}$
circle in the brane worldvolume goes to infinity when we approach the boundary.
Thus, it is not clear if we really have a spontaneously broken global symmetry
as described above. To check this we look for a mode of the  fluctuation of the
$B_{\mu\nu}$ field that corresponds to a massless mode in the dual field theory,
and which is normalizable.

Consider the fluctuation modes of the self--dual two-form field $B$ living on the $M5$-brane.
Recall first  that the  induced metric on the  $M5$-brane is (see (\ref{TheMetric})) given by
\bea\label{IndM1}
ds_{ind}^2=H^{-1/3}\left[
dx^2_{\mu}+(1 + H (|\pa v|^2 + |\pa w|^2)) dz d{\bar z}
\right].
\eea
Classically, the self--dual $B$-field living on the $M5$-brane vanishes.
To study a candidate for a Nambu--Goldstone mode, we only excite the
components of the $B$-field which give rise to scalars in spacetime:
\bea\label{ScalarB}
B=ib~dzd{\bar z}+(\d_\mu \phi dx^\mu dz+{\rm c.c})+{\hat B}_{\mu\nu}
dx^\mu dx^\nu\,.
\eea
All fields appearing here are functions of $(x_\mu,z,\zb)$.
The field $\phi$ can be removed by a gauge transformation, and then the self--duality condition determines ${\hat B}_{\mu\nu}$ in terms of $b$. Focusing on the first term in
(\ref{ScalarB}), we find
\bea
&&B=ib~dzd{\bar z},\quad dB=i(\d_\mu b)dx^\mu dz d{\bar z},\quad
*dB=ig^{z{\bar z}} H^{-1/3} (*_4 d_4b),\nonumber\\
&&d(*dB)=ig^{z{\bar z}}H^{-1/3}\d_\mu\d^\mu b~d^4 x+id_2[(g^{z{\bar z}}H^{-1/3})
(*_4 d_4b)].
\eea
Since the $B$-field is self--dual, the expression in the second line must vanish.
This implies that $b$ is a massless mode, namely, $\pa^\mu\pa_\mu b=0$. Let us check
whether such a mode is normalizable. The second term in the last equation allows us
to determine the dependence of $b$ on $(z,\bar z)$:
\bea
b=H^{1/3}g_{z{\bar z}}{\tilde b}(x_{\mu})=\left[1+{H\over {\lambda_p^2}} (v{\bar v}+w{\bar w})\right]{\tilde b}(x_{\mu}).
\eea
To determine the norm of this expression, we should evaluate
\bea \label{bnorm}
&&\int dz d{\bar z}\sqrt{-g}(-g^{tt})(g^{z{\bar z}})^2(\d_\mu b)^2\sim
\int dz d{\bar z} (\d_\mu {\tilde b})^2,
%
\eea
which clearly diverges.
The expression (\ref{bnorm}) indicates that, as we found in the previous subsections, the
spectrum of this mode is continuous and there is no normalizable four dimensional mode (but
just a continuum corresponding to a massless field in six dimensions). Thus, we cannot view
the global symmetry discussed above as spontaneously broken, and presumably it is not a good
global symmetry of the four dimensional theory for the reasons discussed above. The absence
of Nambu-Goldstone bosons was also noted in a similar situation in \cite{VanRaamsdonk:2009gh},
in a background which also exhibits a continuous spectrum. Note that in the calculation
described above we did not explicitly take into account the self--duality constraint. However,
we have verified that a careful analysis using the action for a self--dual field presented in
\cite{Pasti:1997gx} leads to the same result.

\section{Energy scales}\label{QCDstring}

In this section we will discuss the energy scales that enter the dynamics
of the brane configurations of section 2. We start with the configuration of figure \ref{puresym}
in weakly coupled type IIA string theory, with $\xi$, $\lambda_p$ large in string units. One way
to determine the confinement scale in this theory is to calculate the potential between a heavy
quark and anti-quark separated in $\IR^3$ by a large distance $L$ (not to be confused with the
distance in $x_6$ between the $NS5$-branes, denoted by $L$ in section 2). In QCD, this
potential goes at large distances like $V(L)\simeq LT_{\rm conf}$, where $T_{\rm conf}$ is
the tension of the QCD string, and the masses of glueballs are of order $\sqrt{T_{\rm conf}}$.

In MQCD, this confining string was discussed in \cite{Witten:1997ep}. Although this paper
considered the case of large $g_s$, the construction described in it generalizes trivially to
the weakly coupled theory studied here. The confining string can be viewed as an $M2$-brane
ending on the $M5$-brane described by (\ref{WittProf}). At fixed $z$, this $M5$-brane looks
like $p$ points on a circle  in the $v$ plane, and $p$ points in the $w$ plane. The locations of
the points in the two planes are correlated and change with $z$. The usual type IIA string
is an $M2$-brane wrapped around the $x_{11}$ circle, but if this $M2$-brane ends on the
$M5$-brane, it can be continuously deformed to an $M2$-brane which sits at a fixed value of
$x_{11}$, and stretches between two adjacent points in $v$ and $w$. This
string minimizes its energy if it sits at $x_6=0$, where its tension is
\cite{Witten:1997ep} (up to numerical constants)
\bea\label{ttt}
T_{\rm conf}\simeq {\xi\over pg_sl_s^3}={\xi\over\lambda_pl_s^2} \,.
\eea
From the point of view of the brane configuration of figure \ref{puresym},
this confining string (being an $M2$-brane localized in $x_{11}$)
is really a $D2$-brane stretched between the $D4$-branes.

To have a regime in which the dynamics of the theory is dominated by the confining
string,
the tension (\ref{ttt}) must be well below the fundamental type IIA string scale. This is the
case if $\xi\ll\lambda_p$.
Note that small fluctuations of the fivebrane, of the sort studied in section
\ref{Tfotbaths}, do not give rise to any normalizable states in flat spacetime, and it is not clear
if any discrete states exist.

One may think that when a D-string becomes lighter than the
fundamental string, perturbation theory would break down. However, in our case
this D-string only exists when it is bound to the fivebrane (which we
view as a probe), so it does not affect the validity of perturbation theory
in the bulk (or of the probe approximation to the fivebrane physics).

So far, we discussed the system of section 2 for $N=0$. We now add to the brane
configuration $N\gg p$ infinite $D4$-branes, and restrict to their near-horizon geometry.
Without the curved fivebrane (\ref{WittProf}), the quark-anti-quark potential now takes the
form $V(L) \propto -\lambda_N / L^2$ \cite{Brandhuber:1998er}. This is
obtained by considering the minimal energy type IIA string ending at two points on the
boundary separated by a distance $L$ in $\IR^3$.

When we add the probe fivebrane (\ref{WittProf}), the quark-anti-quark potential can in principle
change. For very small $L$, the type IIA string does not reach the curved fivebrane (\ref{WittProf})
and cannot connect to it, while for very large $L$ the string can go deep into the region of small
$v,w$ where its tension goes to zero, and this gives the minimal energy configuration.
However, when the minimal radial position of the type IIA string is of order $\xi$,
the confining string could have a smaller energy. Clearly, this can only happen if
the tension of the confining string is smaller than the tension of a type IIA string at $\xi$.

\begin{figure}[!]
\begin{center}
\includegraphics[width= 100mm]{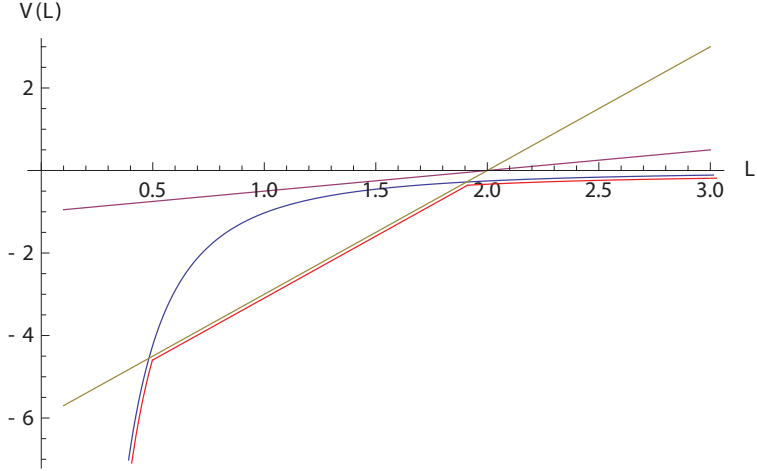}\end{center}
\caption{The quark anti-quark potential $\frac{V(L)\lambda_p^2}{\lambda_N}$ as a function of the distance $\frac{L}{\lambda_p}$.
The blue line denotes the potential
for a string unattached to the fivebrane, while the purple and green
lines are the potentials for attached strings with $q=0.5, 3$,
respectively. For $q=3$, the actual potential is the minimum between
the blue line and the green line, which is drawn in red.
\label{qcdstring}}
\end{figure}

The tension of the confining string (\ref{ttt}) is not affected by
the $D4$-brane background. However, the fundamental type IIA string tension is corrected.
When the $N$ $D4$-branes are at $v=w=0$, one has (for $x_7=0$)
\bea\label{hhhh}
H={g_sNl_s^3\over(|v|^2+|w|^2)^{3\over2}}\,.
\eea
This harmonic function renormalizes the tension of
a type IIA string sitting at some $(v,w)$ to
\bea\label{twoaa}
T_{IIA}={H^{-1/2}\over l_s^2}\,.
\eea
Thus, the ratio between the tensions of the confining string and of a type IIA
string at $u\simeq \xi$ is
\bea\label{ratiohhh}
{T_{\rm conf}\over T_{IIA}}\simeq
\left(\lambda_N l_s^2\over\xi\lambda_p^2\right)^\half\,.
\eea
If
\bea\label{condxx}
{\xi\over l_s^2}\gg{\lambda_N\over\lambda_p^2},
\eea
or equivalently $q \gg 1$ in the notations of the previous section,
there is a range of $L$'s for which the confining string dominates the
quark-antiquark potential. In this range of $L$'s the quark-anti-quark
potential is given by $V(L) = T_{\rm conf} L - c \xi / l_s^2$ for some positive
constant $c$, where the second term comes from the renormalized
energy of the type IIA string going down to the $M5$-brane at $\xi$
and then coming back out. The precise quark-anti-quark
potential is given (see figure \ref{qcdstring}) by the minimum between this expression and $V(L) = -\lambda_N / L^2$.
For $q \ll 1$ the latter expression is always
smaller, but for $q \gg 1$ there is a range of distances where the confining string
dominates the quark-anti-quark force.

In section \ref{Tfotbaths} we saw that the spectrum of mesons in holographic MQCD is
particularly rich in the opposite limit, $q\ll 1$. The masses of the mesons range in
this limit between $\sqrt{q}/\lambda_p$ and $1/\lambda_p$ (see figure \ref{spectrumx7}).
In the regime we consider they are all well below $m_s$. However, since
the fundamental string tension is renormalized (\ref{twoaa}), in order for the mesons
of section  \ref{Tfotbaths} to be well separated from the string excitations, they
must be much lighter than the local string scale at $u\simeq \xi$. One can check that
for the light mesons (\ie\ those with $m \sim \sqrt{q}/\lambda_p$) this follows from (\ref{xinearh}),
while for the heavy ones
(those with $m \sim 1/\lambda_p$) it leads to the requirement
\bea\label{condqq}
q\gg\left(\lambda_p\over\lambda_N\right)^{2\over3}=\left(p\over N\right)^{2\over3}.
\eea
This condition is stronger, but it still allows $q$ to be very small.


\section{Finite temperature}\label{finiteT}


Holography relates the $4+1$ dimensional theory of $N$ $D4$-branes at finite
temperature to string theory in the near-horizon geometry of Euclidean non-extremal
fourbranes \cite{Witten:1998zw}. This geometry is given by\footnote{As before,
the type IIA geometry is obtained by reducing on the $x_{11}$ circle.}
\bea\label{finTMetric}
ds^2&=&H^{-1/3}[fd\tau^2+dx_{i}^2+dx_6^2+dx_{11}^2]+
H^{2/3}[(f^{-1}-1)dr^2+|dv|^2+|dw|^2+dx_7^2],\nonumber\\
C_6&=&H^{-1}d^4x\wedge dx_6\wedge dx_{11}.
\eea
Here we define $r^2 = |v|^2+|w|^2+x_7^2$, which obeys $r=u$ at $x_7=0$ (where all our branes from
here on will be localized).
In addition to the harmonic function $H$ defined in (\ref{TheMetric}), which in this limit takes the form
\bea
H=\frac{u_0^3}{r^3},\qquad u_0^3\equiv \pi\la_Nl_s^2,
\eea
this metric contains a non--extremality factor $f$, which depends on the temperature $T$,
\bea
f=1-\frac{u_T^3}{r^3},\qquad u_T=\frac{16\pi^2}{9}(Tl_s)^2\lambda_N.
\eea
The Euclidean time $\tau=it$, is periodically identified,
\bea
\tau\sim \tau+\frac{1}{T}.
\eea
The relation between $u_T$ and the temperature ensures the smoothness of the
metric (\ref{finTMetric}) at $u=u_T$.

To study the system of section 2 at finite temperature we need to place the $NS$, $NS'$
and $p$ $D4$-branes in the background (\ref{finTMetric}). Since these branes are treated
as probes, they do not change the background. One can think of (\ref{finTMetric}) as
providing a thermal bath in which the four dimensional confining gauge theory is placed.
Even though the bath is five dimensional while the theory we are interested in is
four dimensional, it is clear that $T$ above is also the temperature felt by the four
dimensional degrees of freedom, which are in thermal equilibrium with the five dimensional
ones.\footnote{Similar issues arise in the Sakai-Sugimoto model and
related models, see, e.g., \cite{Aharony:2006da,Parnachev:2006dn,Antonyan:2006qy}.}

Since the geometry is modified by the temperature, we need to determine the shape of the
curved fivebrane in the new geometry.  The free energy is given (in the probe approximation)
by the fivebrane action for this shape (times the temperature).
Recall that at zero temperature, the profile (\ref{WittProf})
has a $U(1)$ symmetry, (\ref{branesym}), which corresponds to a $U(1)$ global symmetry
in the gauge theory of section \ref{fieldth}. From the field theory point of view, it is clear that
this symmetry remains unbroken at finite temperature. In the geometry (\ref{finTMetric}) this
is the statement that the function $f$ is invariant under (\ref{branesym}).
The most general ansatz consistent with this $U(1)$ is given by (\ref{nsansatz}).
Plugging it into the fivebrane Lagrangian (for $x_7=0$) leads to
\bea\label{PRact}
L=H^{-1}\sqrt{f(1+H(u{\dot\phi})^2)\left(1+\frac{H}{f}(u')^2+H(u\alpha')^2
\right)}-H^{-1},
\eea
which is the finite temperature counterpart of the Lagrangian (\ref{Laguncomp}). As in the
zero temperature case, the equations of motion imply that $\phi=x_{11}/\la_p$, and
translation invariance in $\alpha$ and $x_6$ leads to the conserved charges
\bea\label{cccFinTemp}
J&=&\frac{u^2\alpha'\sqrt{f(1+Hu^2/\la_p^2)}}{\sqrt{1+\frac{H}{f}(u')^2+H(u\alpha')^2}}~,\nonumber\\
E&=&H^{-1}-\frac{H^{-1}\sqrt{f(1+Hu^2/\la_p^2)}}{\sqrt{1+\frac{H}{f}(u')^2+H(u\alpha')^2}}~.
\eea
To study the branes at finite temperature it is convenient to invert the relation $u(x_6)$ and
use $u$, rather than $x_6$, as an independent variable. After some algebra, equation
(\ref{cccFinTemp}) can be rewritten as
\bea
\label{alPrime}
\frac{d\alpha}{du}&=&\frac{J}{u^2}\frac{dx_6}{du}\frac{1}{1-EH}~,\\
\label{Xprime}
\frac{dx_6}{du}&=&\pm\frac{u}{J\sqrt{f}}(1-EH)\left[\frac{u^2}{J^2}f(H^{-1}+\frac{u^2}{\la_p^2})-1-
\frac{u^2}{J^2}(1-EH)^2H^{-1}\right]^{-1/2}~.
\eea
It is convenient to define
\bea\label{defG}
G(u)=\frac{u^2}{J^2}f(H^{-1}+\frac{u^2}{\la_p^2})-1-
\frac{u^2}{J^2}(1-EH)^2H^{-1}~,
\eea
in terms of which (\ref{alPrime}), (\ref{Xprime}) take the form
\bea
\label{newPrime}
\frac{d\alpha}{du}&=&\pm\frac{1}{u\sqrt{fG}}~,\\
\label{newXprime}
\frac{dx_6}{du}&=&\pm\frac{u}{J\sqrt{fG}}(1-EH)~.
\eea
Note that $x_6$ is not a single--valued function of $u$, and we have to consider two branches, which correspond to the two signs in (\ref{newPrime}), (\ref{newXprime}). These branches connect at the point
where $u$ reaches its minimal value $u_{min}$, which is determined by the condition
\bea\label{condtwo}
G(u_{min})=0.
\eea
The integrals of motion, $E$ and $J$, are determined by imposing appropriate boundary conditions on
$x_6$ and $\alpha$. As discussed in section 2.1, at large $u$ the distance between the $NS5$-branes
goes to infinity, so to define the theory one has to introduce a cutoff  $u_\infty$ and to impose
boundary conditions at $u=u_\infty$.

The boundary has two components, corresponding to the $NS$ and $NS'$-branes (or, equivalently,
negative and positive branches of (\ref{newPrime}), (\ref{newXprime})). On the negative ($NS$) branch,
the boundary conditions are $\alpha(u_\infty)=\epsilon$, $x_6(u_\infty)=-L/2$; on the
positive one they are $\alpha(u_\infty)=\pi/2-\epsilon$, $x_6(u_\infty)=L/2$, where $\epsilon \to 0$
as $u_{\infty} \to \infty$. The boundary conditions
preserve the symmetry of the equations of motion $x_6\to -x_6$; thus, $\alpha(u_{min})=\pi/4$.

It is important to emphasize that $u_\infty$, $L$, $\epsilon$ above are independent of temperature
-- the boundary conditions are used to define the theory at the UV cutoff scale, and are independent of
the state.\footnote{Of course, we must choose the UV cutoff to be sufficiently large,
$u_\infty\gg u_T$.}
In particular, $\epsilon$ can be calculated in the zero temperature theory by using (\ref{NSUsol}).

The constants $J$ and $E$ (\ref{cccFinTemp}) can be calculated as a function of temperature
by integrating the equations of motion. Consider first the integral of (\ref{newPrime}). In principle we
should integrate from $u_{\rm min}$ to $u_\infty$, but the resulting integral is convergent, so one
can send the upper limit of integration to infinity (and thus $\epsilon\to 0$). This leads to
\bea\label{DtermP}
\int_{u_{min}}^\infty\frac{du}{u\sqrt{fG}}=\frac{\pi}{4}.
\eea
This, together with (\ref{condtwo}), gives one condition on $J$, $E$.

The second condition comes from integrating (\ref{newXprime}). This integral is divergent at large
$u$,  so we need to be more careful with it. Recall that at zero temperature the theory is characterized
by the ``QCD scale'' $\xi$, which enters the relation (\ref{lcutoff}) between $L$ and $u_\infty$.
Denoting the integration constant related to $\xi$ via  (\ref{XiVsJ}) by $J_0$, and the corresponding
function $G$ (\ref{defG}) by $G_0$,\footnote{So $G_0=(u^2/ J_0\lambda_p)^2-1$.} we can write
(\ref{lcutoff}) as
\bea\label{zeroT}
\frac{L}{2}=\int_{\sqrt{J_0\la_p}}^{u_\infty}\frac{udu}{J_0\sqrt{G_0}}.
\eea
Integrating the finite temperature equation of motion (\ref{newXprime}) leads to the relation
\bea\label{condone}
\frac{L}{2}=\int_{u_{min}}^{u_\infty}\frac{udu}{J\sqrt{fG}}(1-EH)=
\int_{\sqrt{J_0\lambda_p}}^{u_\infty}\frac{udu}{J_0\sqrt{G_0}}.
\eea
To remove the UV cutoff, it is useful to rewrite (\ref{condone}) such that the limit
$u_\infty\to\infty$ is smooth. This can be done by subtracting the two integrals
in (\ref{condone}) and combining them. The resulting integral is finite
in the limit $u_\infty\to\infty$. In this limit one finds an integral  equation in
which $u_\infty$ and $L$ have been traded for the physical (``QCD'') scale
$\xi$. This is the brane analog of the process of renormalization in QCD.

Equations (\ref{condtwo}), (\ref{DtermP}) and (\ref{condone}) determine $E$ and $J$ as
functions of $J_0$ (or $\xi$) and the temperature.  The profile of the brane is then
determined by solving (\ref{newPrime}) and (\ref{newXprime}). The free energy of the
solution (divided by the temperature) is given by
\bea\label{ConnEnrgy}
{\cal E}^{(con)}&=&
\int_{u_{min}}^{u_\infty} du H^{-1}\left[\sqrt{f(1+\frac{u^2H}{\la_p^2})\left((\d_ux_6)^2+\frac{H}{f}+H(u\d_u\alpha)^2
\right)}-H^{-1}\d_u x_6 \right]\nonumber\\
&=&\int_{u_{min}}^{u_\infty} \frac{udu}{\sqrt{fJ^2 G}}\left[H^{-1}(f-1)+\frac{u^2}{\la_p^2} f+E\right],
\eea
where we restricted to the positive branch; therefore the full free energy of the
curved fivebrane is $2{\cal E}^{(con)}$.

The connected solution which has been discussed so far, corresponds to a confining vacuum of
the gauge theory of section 3 at finite temperature. Another solution of the equations of motion, which
corresponds to the Higgs branch, is a configuration of two disconnected fivebranes which descend
from large values of $u$ to the horizon located at $u=u_T$. Looking at equation (\ref{Xprime}) and requiring $x_6'$ to be real for all $u>u_T$, we find that such a disconnected solution must have
\bea
J=0,\qquad E=\frac{u_T^3}{u^3_0}.
\eea
For these values of parameters equation (\ref{newXprime}) simplifies
(we only look at one of the branches)
\bea
\frac{dx_6}{du}=\left(\frac{u^2}{\la_p^2}+\frac{u_T^3}{u^3_0}\right)^{-\half}.
\eea
It has a unique solution satisfying the relation (\ref{zeroT}) between $L$ and $u_\infty$,
\bea\label{DisconSoln}
x_6=\la_p\ln \left(u+\sqrt{u^2+\frac{\la_p^2u_T^3}{u^3_0}}\right)-\frac{\la_p}{2}\ln (2J_0\lambda_p).
\eea
This solution should only be considered in the exterior of the black hole, $u>u_T$.
The derivative $\frac{d\alpha}{du}$ vanishes in this region (see (\ref{alPrime})); the
solution has $\alpha=\pi/2$ and describes an $NS'$-brane with $p$ $D4$-branes attached.

At zero temperature the connected and disconnected solutions have the same energy
(in the limit when the cutoff $u_\infty$ is sent to infinity), in agreement with the fact that
they describe two supersymmetric vacua of the same theory. At finite temperature, one
of them can have lower free energy. The free energy of the connected solution is given by
twice (\ref{ConnEnrgy}); for the disconnected one we find a free energy $2{\cal E}^{(dis)}$ with 
\bea\label{DisconEnergy}
{\cal E}^{(dis)}&=&\int_{u_T}^{u_\infty} du\left[\sqrt{f(H^{-1}+\frac{u^2}{\la_p^2})\left(H^{-1}(\d_ux_6)^2+\frac{1}{f}
\right)}-H^{-1}\d_u x_6\right]\nonumber\\
&=&\frac{1}{2}\left[u\sqrt{\frac{u^2}{\la_p^2}+\frac{u_T^3}{u^3_0}}-\la_p\frac{u_T^3}{u^3_0}
\ln\left\{u+\sqrt{u^2+\la_p^2\frac{u_T^3}{u^3_0}}\right\}\right]_{u_T}^{u_\infty}.
\eea

Let us demonstrate that at small but finite value of $u_T$ this free energy is smaller than (\ref{ConnEnrgy}). Near $u_T=0$, (\ref{DisconEnergy}) behaves as
\bea\label{DiscExpand}
{\cal E}^{(dis)}
&=&\frac{1}{2\la_p}\left[u_\infty^2+\frac{\la_p^2u_T^3}{2u^3_0}(1+\ln\left(\frac{u^2_T}{u^2_\infty}\right)) -u_T^2+o(u_T^3)\right].
\eea
This should be compared with a similar expansion of (\ref{ConnEnrgy}),
\bea\label{IntermConn}
{\cal E}^{(con)}
&=&\int_{u_{min}}^{u_\infty} \frac{udu}{\sqrt{fJ^2 G}}\left[\frac{u^2}{\la_p^2}+E\right]-
u_T^3\int_{u_{min}}^{u_\infty}
\frac{udu}{u^3\sqrt{fJ^2 G}}\left[H^{-1}+\frac{u^2}{\la_p^2}\right]\\
&=&\frac{1}{\la_p^2}\int_{u_{min}}^{u_\infty} \frac{u^3du}{\sqrt{fJ^2 G}}-
u_T^3\int_{\sqrt{J_0\la_p}}^{u_\infty} \frac{du}{u^2\sqrt{u^4/\la_p^2-J_0^2}}\left[H^{-1}+\frac{u^2}{\la_p^2}-\frac{E}{u_T^3}u^3\right]+
o(u_T^3).\nonumber
\eea
Here we used the fact that the ratio $E/u_T^3$ remains finite as $u_T$ goes to zero (this can be
shown by performing a small $u_T$ expansion of (\ref{DtermP}) and (\ref{condone})) as well as
(\ref{defG}). To simplify the first term in (\ref{IntermConn}), we observe that
\bea
fJ^2 G=\frac{u^4}{\la_p^2}-J^2+2Eu^2+\frac{u_T^3}{u^3}\left[J^2-H^{-1}u^2-2\frac{u^4}{\la_p^2}\right]+
o(u_T^3).
\eea
This equation leads to an expression for $du^4$ in terms of $d(fJ^2 G)$, and substituting the result into the first term in (\ref{IntermConn}), we find
\bea\label{IntCommExtr}
&&\frac{1}{4\la_p^2}\int_{u_{min}}^{u_\infty} \frac{du^4}{\sqrt{fJ^2 G}}\nonumber\\
&&\quad=\frac{1}{4}\left[
\left.2\sqrt{fJ^2G}\right|_{u_{min}}^{u_\infty}-u_T^3\int_{u_{min}}^{u_\infty} \frac{1}{\sqrt{fJ^2 G}}
d\left\{\frac{2E}{u_T^3}u^2+\frac{J^2}{u^3}-\frac{1}{uH}-2\frac{u}{\la_p^2}\right\}+o(u_T^3)
\right]\nonumber\\
&&\quad\sim\frac{u_\infty^2}{4\la_p}\left(2-\frac{\la^2_p}{u_\infty^2}\frac{u_T^3}{u_0^3}+
2E\frac{\la_p^2}{u_\infty^2}\right)\nonumber\\
&&\qquad-\frac{u_T^3}{4}\int_{\sqrt{J_0\la_p}}^{u_\infty} \frac{du}{\sqrt{u^4/\la_p^2-J_0^2}}
\left\{\frac{4E}{u_T^3}u-\frac{3J^2}{u^4}-\frac{2u}{u_0^3}-\frac{2}{\la_p^2}\right\}+o(u_T^3).
\eea
At the last stage we dropped some terms which vanish in the limit $u_\infty\rightarrow\infty$. Substituting (\ref{IntCommExtr}) into (\ref{IntermConn}) and performing the integral in the term
that diverges at large $u_\infty$, we find
\bea
{\cal E}^{(con)}&=&\frac{u_\infty^2}{2\la_p}+u_T^3\left[
\frac{\la_p E}{2u_T^3}-\frac{\la_p}{4u_0^3}\right]+
\frac{\la_p u_T^3}{4u_0^3}\ln\left(\frac{J_0\lambda_p}{u_\infty^2}\right)\nonumber\\
&+&
\frac{u_T^3}{4}\int_{\sqrt{J_0\la_p}}^{u_\infty} \frac{du}{\sqrt{u^4/\la_p^2-J_0^2}}
\left\{\frac{3J^2}{u^4}-\frac{2}{\la_p^2}\right\}+o(u_T^3)
\eea
For small values of the nonextremality parameter $u_T$, this expression is larger than
(\ref{DiscExpand}),
\bea
{\cal E}^{(con)}-{\cal E}^{(dis)}=\frac{u_T^2}{2\la_p}+
\frac{\la_p u_T^3}{4u^3_0}\ln\left(\frac{J_0\lambda_p}{u^2_T}\right)+O(u_T^3).
\eea
Thus, we see that the disconnected configuration is thermodynamically preferred for all non-zero $T$,
and the phase transition from the confining to the Higgs phase (if we start from the confining
phase at zero temperature) occurs at $T=0$. Of course,
we can still study the confining phase at (sufficiently small) non-zero $T$, but it is meta-stable
in this regime.

The above discussion is  natural from the field theory point of view. The Higgs phase of the
gauge theory discussed in section \ref{fieldth} (corresponding to the brane configuration of
figure \ref{fullhiggs}) has $(N+p)^2$ massless fields, while the confining phase has $N^2$.
Since the former has more massless degrees of freedom, its free energy is lower; the
difference is an order $p/N$ effect. Our result on the energetics of the branes implies that this
behavior persists at strong coupling.

It is natural to ask whether it is possible to deform the brane system so as to shift the
deconfinement phase transition to finite temperature. One way to do that is to start at
zero temperature with a brane configuration in which the $NS5$-branes are displaced
relative to the $N$ $D4$-branes in the $x_7$ direction by the distance $X_7\ne 0$.
The free energies of the confining (figure \ref{massive}) and Higgs (figure \ref{fullhiggs})
branches depend on $X_7$, and it is possible that the transition between them occurs
at finite $T$.

A complication in this analysis is that at finite temperature the curved connected and
disconnected fivebranes that correspond to the two branches are no longer located at
fixed $x_7$. One can show that they develop a profile in $x_7$ that (at large $x_6$) changes
logarithmically with $u$. This bending can be understood from field theory as due
to the fact that in order to have a non-zero vacuum expectation value of $x_7$ at finite temperature, one has to
add to the Lagrangian a $4+1$ dimensional FI D-term for the  $U(1)$'s on the semi-infinite
$D4$-branes. The presence of this logarithmic mode complicates the analysis of
the energetics, and we will leave it to future work.

\section{ Non-supersymmetric generalizations}\label{nonsusygen}

So far we have discussed brane configurations with $d=4$ ${\cal N}=1$
supersymmetry. There are many possible generalizations, to different
dimensions and different amounts of supersymmetry. In this section
we discuss two non-supersymmetric brane systems that are closely related
to those studied in this paper. The first is a configuration similar to
that discussed in section \ref{holemb} and drawn in figure \ref{origin},
but with the two $NS5$-branes taken to be parallel and having opposite
orientations (so that in the classical limit they look like an $NS$-brane
and an $\overline{NS}$-brane). We analyze this model both at zero and at finite
temperature. The second system is similar to the model discussed in section
\ref{compcirc}, with a compact $x_6$ direction (see figure \ref{compactified}),
but with anti-periodic boundary conditions for the fermions around the circle.

\subsection{ The $NS-\overline{NS}$ system}

We start with the brane configuration of figure \ref{origin}, with the
$NS'$-brane replaced by a second $NS5$-brane parallel to the first one,
but with opposite orientation; we will refer to it as an $\overline{NS}$-brane.
In the full string theory, this configuration is unstable to gravitational
attraction of the fivebranes, but in the decoupled theory of the $N$
$D4$-branes, which we focus on here, the mode governing the separation
between the fivebranes is non-normalizable, and this instability is absent.

The field theory in this case is similar to the one discussed in section
\ref{fieldth} for the configuration of figure \ref{origin}. In that case
the couplings of the fundamental fields living at the brane intersections
to the adjoint fields preserved $\cN=2$ supersymmetry at each intersection
separately, but only $\cN=1$ for both intersections together, while in this
case there are no common supersymmetries. Classically the brane configuration
has a moduli space corresponding to moving the $D4$-branes on the interval
along the $NS5$-branes; quantum mechanically, we expect this moduli space
to be lifted. It would be interesting to see whether at weak coupling the
field theory potential on this moduli space is attractive or repulsive.

In the string dual, this brane configuration (for $x_7=0$) is described by the Lagrangian
(\ref{Laguncomp}) with $\a'=0$.  Expressing the Lagrangian  in terms
of dimensionless coordinates
\be\label{rescale}
u\rightarrow \frac{u}{\hat u},\quad x_6 \rightarrow \frac{x_6}{\lambda_p},\qquad
\hat u \equiv\frac{\pi \lambda_N l_s^2}{\lambda_p^2},
\ee
as was done above equation (\ref{defq}), we find (up to an overall constant)
\be\label{Lrescaled}
 {\cal L}=u^{5/2} \sqrt{( 1  + u)\left( 1 +  \frac{ (u')^2}{ u^3}\right)} -  u^3.
\ee
The corresponding ``conserved Hamiltonian"  is given by
\be
E=u^3- \frac{u^{5/2} \sqrt{( 1 + u)}}{\sqrt{ 1 +  \frac{ (u')^2}{ u^3}}}.
\ee
This implies the following equation
\be\label{uPrimeT0}
u'=\pm u^{3/2}\sqrt{\frac{u^5(1+u)}{(u^3-E)^2}-1}.
\ee
It is easy to see that for $E=0$ there is no connected solution but  only a disconnected one, of the form
(\ref{discU}) $u = K \exp\left(\pm x_6/\lambda_p\right)$, describing a configuration in which the gauge
symmetry is broken as in section \ref{fieldth}. In the discussion of section \ref{setups}, the
connected solution also had $E=0$ due to supersymmetry, but now with no supersymmetry there is no
reason why a connected solution should have this property. Connected solutions exist for any $E < 0$, and
they are characterized by having a minimum point where $u'=0$. This point is the maximal solution $u=u_*$
to the equation for the vanishing of $u'$,
\be\label{u0T0}
u_*^5 + 2E u_*^3 -E^2=0.
\ee
It is easy to check (this
is a special case of a computation done in the next subsection) that the connected configuration always has
lower energy than the disconnected one, so it describes the vacuum of this non-supersymmetric field
theory. We did not analyse explicitly the stability of this brane configuration, but we expect it to be stable,
and to otherwise exhibit similar physics to what we found in the previous sections.

The theory is parametrized by $u_*$ (the minimal value of $u$), which is an analog of the ``QCD scale'' $\xi$
introduced in section \ref{setups}. Integrating (\ref{uPrimeT0}) between $u_*$ and the UV cutoff $u_\infty$, we find
an analog of the relation (\ref{lcutoff}) between the QCD scale and the distance between the branes at
$u=u_\infty$:
\bea\label{LAsIntegr}
\frac{L}{2}=\int_{u_*}^{u_\infty}\frac{(u^3-E)du}{\sqrt{u^3(u^5+2E u^3-E^2)}}.
\eea
Here $E(u_{*})$ is a negative number that solves equation (\ref{u0T0}). The integral on the right--hand
side of (\ref{LAsIntegr}) diverges logarithmically for large $u_\infty$, so it is convenient to define a
renormalized quantity,
\bea\label{LRen}
L_{ren}=\lim_{u_\infty\rightarrow\infty}(L-2\ln(u_\infty))=2\int_{u_*}^{\infty}\left[\frac{(u^3-E)}{\sqrt{u^3(u^5+2E u^3-E^2)}}-\frac{1}{u}\right] du -2\ln(u_*),
\eea
which remains finite as $u_\infty\to\infty$. Inverting this relation, one finds $u_*$ as a function of $L_{ren}$, which is shown
in the red line in figure \ref{UtLcr}.

\subsection{ The $NS-\overline{NS}$ system at finite temperature}

At finite temperature,
the $NS-\overline{NS}$ system is described by the Lagrangian (\ref{PRact}) with $\alpha'=0$. Performing the re-scaling (\ref{rescale}) in (\ref{PRact}),
or equivalently introducing the thermal factor $f(u) = 1 - u_T^3 / u^3$ into  the Lagrangian (\ref {Lrescaled}), we find\footnote{Notice that this Lagrangian differs from (\ref{PRact}) by an overall factor, and that $u_T$ here
is rescaled with respect to $u_T$ of section \ref{finiteT}.}
\be
{\cal L}=u^{5/2} \sqrt{f( 1  + u)\left( 1 +  \frac{ (u')^2}{f u^3}\right)} -  u^3.
\ee
The corresponding Hamiltonian for translations in $x_6$ is given by
\be
E_T=u^3- \frac{u^{5/2} \sqrt{f( 1  + u)}}{\sqrt{ 1 + \frac{(u')^2}{f u^3}}}.
\ee
The minimal value of $u$ for the connected configuration, $u=u_{min}$,
corresponds to $u'=0$, which is given by the solution of
\be\label{min}
u_{min}^5 - u_T^3 (  u_{min}^2 + u_{min}^3) + 2E_T u_{min}^3 - E_T^2=0.
\ee
The disconnected configuration has $u'=0$ at $u_{min}=u_T$, where according to (\ref{min}) $E=u_T^3$.

\begin{figure}[!]
\begin{center}
\includegraphics[width= 120mm]{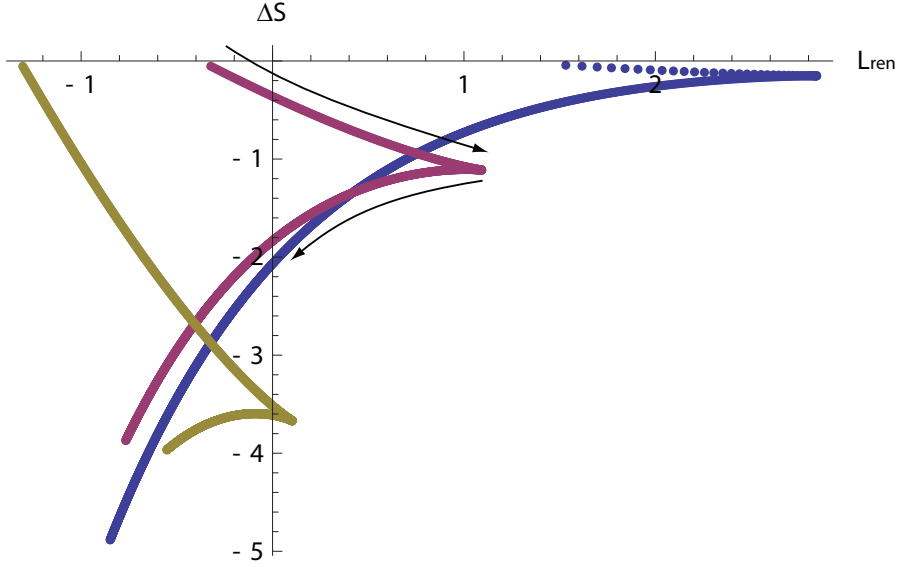}\end{center}
\caption{ $\Delta S$ as a function of $L_{ren}$ for $u_T= 0.5$ (blue), $u_T=1$ (red), and $u_T=1.5$ (yellow). The arrows indicate in which direction the position of the brane $u_{min}$ increases.\label{acdif010308}}
\end{figure}

The connected configuration is characterized by the asymptotic separation $L$, which is
given in terms of $u_{min}$ and $u_T$ as follows
\be\label{Luzero}
L= 2\int_{u_{min}}^{u_{\infty}} \frac{du}{u'} = 2  \int_{u_{min}}^{u_\infty} \frac{u^3-E_T}{u^{3/2}\sqrt{f}}
\frac{1}{\sqrt{u^5  - u_T^3 (  u^2 + u^3) + 2E_T u^3 - E_T^2}} \, du.
\ee
The separation $L$ can be renormalized in the same manner as in (\ref{LRen}).
The action of the connected configuration is given by (up to an overall constant)
\be
S^{(con)}= 2  \int_{u_{min}}^{u_\infty} \frac{u^{3/2}}{\sqrt{f}}\frac{f(u^3 + u^2 ) - u^3+ E_T}{\sqrt{u^5  - u_T^3 (  u^2 + u^3) + 2E_T u^3 - E_T^2}}\, du \,,
\ee
while the action of the disconnected configuration is
\be
S^{(dis)}= 2  \int_{u_T}^{u_\infty}
\frac{u^2 du}{\sqrt{u^2 + u_T^3}}\,.
\ee
The last integral can be evaluated exactly leading to
\be
S^{(dis)}=  \left[ u \sqrt{u^2 + u_T^3} -  u_T^3 \ln [ 2( u+ \sqrt{u^2 + u_T^3})]\right]^{u_\infty}_{u_T}.
\ee
The question of which of the two configurations, the connected one or the disconnected one,
is preferable at any given temperature is determined by the difference of the corresponding
free energies. In our approximation this is translated into the difference between the
classical actions of the connected and disconnected configurations. Each action separately
is divergent as $u_\infty\to\infty$, but the difference between them is finite, and is given by
\bea
\Delta S\equiv S^{(con)}- S^{(dis)} &=&
 2 \int_{u_{min}}^\infty {u^{3/2}}\left\{\frac{1}{\sqrt{f}}\frac{f u^2  - u_T^3+ E_T}{\sqrt{u^5 - u_T^3 ( u^2 + u^3) + 2 E_T u^3 - E_T^2}} - \frac{ u^2 }{\sqrt{u^2  +{u_T^3}}}    \right\} du\CR
 &&\quad -\left[ u \sqrt{u^2 + u_T^3} -  u_T^3 \ln [ 2( u+ \sqrt{u^2 + u_T^3})]\right]^{u_{min}}_{u_T}.
\eea
The difference of the actions as a function of the asymptotic separation distance is shown, for some
 specific values of the temperature, in figure \ref{acdif010308}.

\begin{figure}[!]
\begin{center}
\includegraphics[width= 80mm]{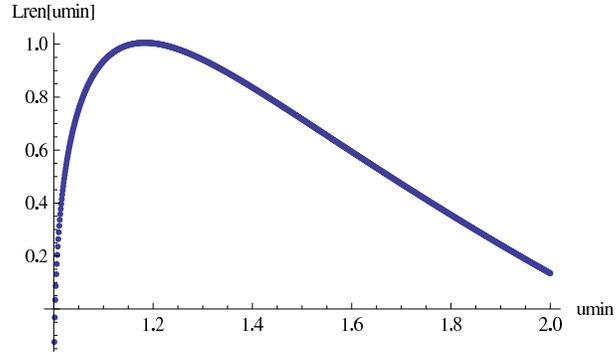}\end{center}
\caption{The asymptotic separation distance  $L_{ren}$ as a function  of $u_{min}$ for $u_T=1$.\label{Lcr}}
\end{figure}
\begin{figure}[!]
\begin{center}
\includegraphics[width= 80mm]{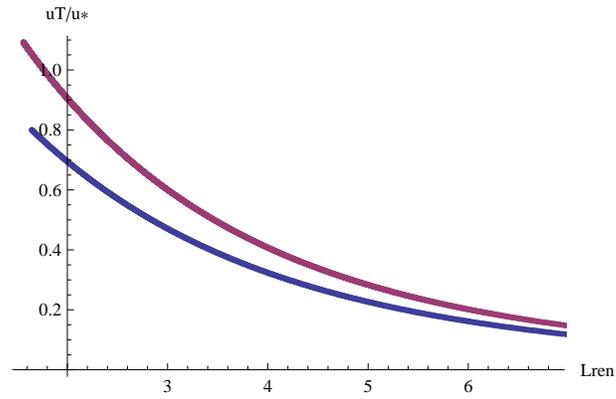}\end{center}
\caption{$u_*$ (red,higher) (at $T=0$), and the phase transition temperature $u_T$ (blue,lower) as a function of the
renormalized distance $L_{ren}$.
 \label{UtLcr}}
\end{figure}

The qualitative dependence of $\Delta S$ on $L$ depicted in the figures is the same for any  temperature $u_T>0$.
From the figures it is evident that $\Delta S$ is always negative, and hence the connected solution is preferred.
However, it is also evident that for any temperature there is a critical asymptotic separation distance $L_{cr}$
(for a fixed UV cutoff), above which the connected configuration no longer exists. This is manifest in figure
\ref{Lcr} which describes the separation distance $L$ as a function of $u_{min}$ for a particular temperature.
Increasing $u_{min}$ from $u_T$ and correspondingly changing $E_T$ according to (\ref{min}), $L$ first increases
to $L_{cr}$, and then decreases. Therefore, for any given separation distance $L$, when the temperature is raised,
eventually $L$ becomes larger than the critical value and hence the connected configuration will cease to exist.
The dependence of the phase transition temperature $u_T$ on the renormalized $L_{cr}$, together with the
 corresponding value of $u_*$ at $T=0$, are shown in figure \ref{UtLcr}.

The precise value of the phase
transition point may be found by solving $\partial L / \partial u_{min}=0$ for $u_{min}$ (at fixed temperature $u_T$), with $L(u_{min})$ given by
(\ref{Luzero}); note that this depends on $u_{min}$ both through $E_T$ and through the lower bound on the integration
region (and that the resulting equation has a finite limit as $u_{\infty} \to \infty$). Inverting the function $u_{min}(u_T)$ that we find gives us the phase transition temperature for a given
value of $u_{min}$ and of $L=L(u_{min})$. At this temperature there will be a first order phase transition to a disconnected configuration.
 We have thus shown that unlike the supersymmetric case discussed in section \ref{finiteT}, for the non-supersymmetric model there is a first order phase transition between the two phases at non-zero temperature.

\subsection{A fivebrane in the cigar topology}

The second non-supersymmetric model we discuss is similar to the compactified model of
section \ref{compcirc}, but with anti-periodic boundary conditions for the fermions.
This system, which breaks supersymmetry, is not stable in flat space, but it is stable
in the near-horizon limit of the $D4$-branes \cite{Witten:1998zw}. The bulk geometry
in this case is a double-Wick-rotation of the near-extremal $D4$-brane solution, and in
this background the $(x_6,u)$ coordinates have the topology of a cigar (as opposed to the
cylinder discussed in section \ref{compcirc}). The eleven dimensional background is given
by (compare to (\ref{finTMetric}))
\bea\label{finTMetricCmp}
ds^2&=&H^{-1/3}[-dt^2+dx_{i}^2+ f(r) dx_6^2+dx_{11}^2]+
H^{2/3}[(f(r)^{-1}-1)dr^2+|dv|^2+|dw|^2+dx_7^2],\nonumber\\
C_6&=&H^{-1}d^4x\wedge dx_6\wedge dx_{11}.
\eea
The function $f$ is given by $f(r)=1-\left( \frac{u_{\Lambda}}{r} \right)^3$, where
$u_\Lambda$ is the value of the radial coordinate $r$ at the tip of the cigar.

\begin{figure}[h]
\begin{center}
\includegraphics[width= 80mm]{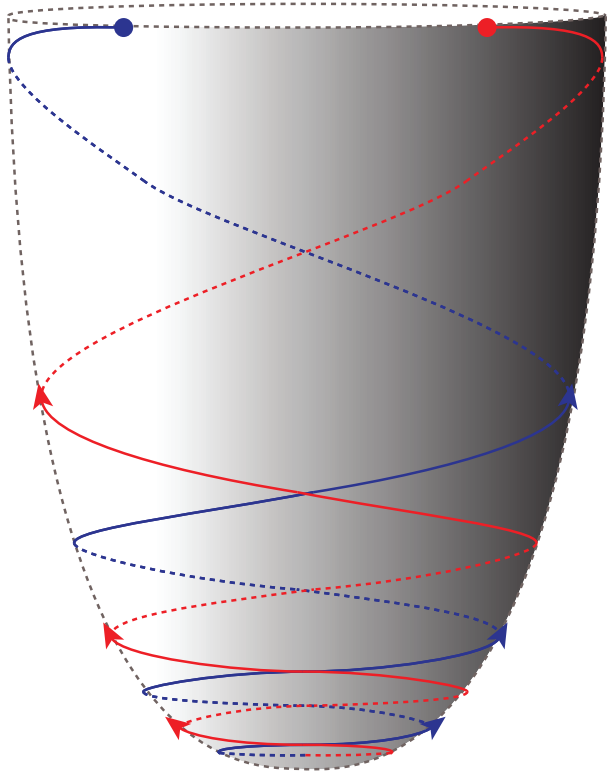}\end{center}
\caption{ The spiraling profile over the cigar background.
Again, we used a red line for the ``downward" brane and a blue one for the climbing one.\label{cigarspiral}}
\end{figure}
Upon reduction to ten dimensional type IIA supergravity, the corresponding metric,
dilaton and RR form read
\begin{eqnarray}\label{S_S_metric}
ds^2\!\!\!\!&=&\!\!\!\bigg( \frac{r}{R_{D4}}\bigg)^{3/2}\!\bigg[\! \!-\!\!dt^2\!+\!\delta_{ij}dx^idx^j+f(r)dx_6^2\bigg]
\!+\!\bigg( \frac{R_{D4}}{r}\bigg)^{3/2}\!\bigg[\frac{dr^2}{f(r)}\!+\!r^2d\Omega_4^2\bigg],\\ \nonumber
F_4\!&=&\!\frac{2\pi N_c}{V_4}\epsilon_4\ \ ,\ \ e^{\phi}=g_s\bigg( \frac{r}{R_{D4}}\bigg)^{3/4}
 , \ R_{D4}^3=u_0^3=\pi g_sN_cl_s^3 .
\end{eqnarray}
Note that the transformation from the extremal background  to the non-extremal one is not just a  change of
the warp factor $H(r)$ which, as we have seen in (\ref{ueqzeroe}), does not modify the brane profile.
Fermions in the cigar background must be taken to be anti-periodic around the $x_6$ circle. Thus,
supersymmetry is broken.

The fivebrane Lagrangian (for $x_7=0$) that follows from (\ref{S_S_metric}),  again  rescaling  (\ref{rescale}), is given by (up to a multiplicative constant)
\be
{\cal L}=  u^{5/2} \sqrt{1+  u}\sqrt{\left(1-\frac{ u_\Lambda^3}{ u^3}\right)+\frac{( u')^2 + ( u \alpha')^2}{ u^3- u_\Lambda^3} } -  u^3,
\ee
The Hamiltonian for translations in $x_6$ is now
\be
E= u^3 -\frac{u^{5/2}(1-\frac{u_\Lambda^3}{u^3})\sqrt{1+ u}}{\sqrt{(1-\frac{u_\Lambda^3}{u^3})+\frac{(u')^2+(u\alpha')^2 }{u^3-u_\Lambda^3} }}\,.
\ee
The conserved charge associated with the shift symmetry of $\alpha$ is
\be
J= \frac{u^{5/2} \sqrt{1+ u}\left (\frac{\alpha'}{u\left(1-\frac{u_\Lambda^3}{u^3}\right)}\right )}{\sqrt{(1-\frac{u_\Lambda^3}{u^3})+\frac{(u')^2+ (u\alpha')^2}{u^3-u_\Lambda^3}}}\,.
\ee
In general it is hard to solve these equations exactly; it is simpler to do this in the
case $E=0$, but this case now does not lead to the correct change in $\alpha$ to connect
the $NS$ and $NS'$-branes. We expect that there should still exist an infinite sequence
of connected solutions, as discussed in section \ref{compcirc} and as depicted in figure
\ref{cigarspiral}, but now these solutions will no longer be degenerate. It would be
interesting to analyze these solutions in detail, to see which solution has the lowest energy.
In the limit $\xi \gg u_{\Lambda}$ we can analyze this by working at leading order in $u_{\Lambda}$;
we find that when the brane lies in a region that the parameter $q$ defined in (\ref{defq}) obeys
$q \ll 1$, the solutions with lower $\xi$ (more cascade steps) have lower energy, so that the minimal
energy configuration has $\xi$ close to $u_{\Lambda}$, while for $q \gg 1$ the situation is the opposite.
Due to the topology of the background it is obvious that in the present case we cannot have a disconnected
profile, since the $NS5$-brane does not have where to end.

\centerline{}
\centerline{\bf Acknowledgements}
\centerline{}

We thank A. Giveon for discussions.
The work of OA, DK, JS and SY was supported in part by the
Israel-U.S. Binational Science Foundation. The work of
OA, JS and SY was supported in part by a  research center
supported by the Israel Science Foundation (grant number 1468/06)
and by a grant (DIP H52) of the German Israel Project Cooperation.
The work of OA was supported in part by the Minerva foundation
with funding from the Federal German Ministry for Education and
Research.  The work of DK was supported in part by DOE grant
DE-FG02-90ER40560 and NSF grant 0529954. The work of OL was supported
in part by NSF grant 0844614. DK thanks the Weizmann Institute and OL
thanks the University of Chicago for hospitality during part of this work.


\appendix
\section{Approximations in the derivation of the spectra of the fluctuations in the transverse directions}

The equations of motion that follow  from the Lagrangian  (\ref{LagrMs11a})
are rather complicated, so to get at least some qualitative understanding on the structure of the bound states we invoke several approximations.
Since the dominant contribution of  $\frac{H_0}{((pR)^2+u^2H)_0}$ comes from  the vicinity of $x_6=0$, where
$V_0\sim W_0$, to get a rough estimate of the eigenvalues we replace
$V_0/u_0$ and $W_0/u_0$ by $\frac{1}{\sqrt{2}}$, and hence
\bea\label{AprAA}
\left[V_0V_1'-W_0W_1'-V_0V_1-W_0W_1\right]^2\approx
\frac{u_0^2}{2}\left[V_1'-W_1'-V_1-W_1\right]^2=
u_0^2\left[V_-'-V_+\right]^2.
\eea
Here we introduced
\bea
V_+=\frac{V_1+W_1}{\sqrt{2}},\qquad V_-=\frac{V_1-W_1}{\sqrt{2}}.
\eea
This approximation leads to the Lagrangian
\bea\label{ApprNov19}
L&\approx&\frac{1}{2}\int dx_6 \left[(V_+')^2+(V_-')^2+
\left(1-(m p R)^2-\frac{(m p R)^2}{q}\frac{1}{\sqrt{2\cosh[2x_6]}}\right)(V_+^2+V_-^2)\right.
\nonumber\\
&&\left.-
\frac{(mpR)^2}{q}\frac{1}{\sqrt{2\cosh[2x_6]}} \left[(mpR)^2+\frac{(mpR)^2}{q}\frac{1}{\sqrt{2\cosh [2x_6]}}\right]^{-1}(V_+-V_-')^2\right]\nonumber\\
&\equiv&\frac{1}{2}\int dx_6\left[(V_+')^2+(V_-')^2+
a(x_6)(V_+^2+V_-^2)+b(x_6)(V_+-V'_-)^2\right].
\eea
This is still a complicated coupled Lagrangian. In the limit of $q<<1$ we may be justified to assume
\bea
-a\sim \frac{(m p R)^2}{q}\frac{1}{\sqrt{2\cosh [2x_6]}}\gg 1,\qquad
b\approx 1,\qquad b'\sim q\ll 1.
\eea
Using this approximation we get a decoupling of $V_+$ and $V_-$ in the Lagrangian  so that for the former we find
\bea\label{WKBlagrtwo}
L&\approx&\frac{1}{2}\int dx_6\left[(V_+')^2+
\left(1-(m p R)^2-\frac{(m p R)^2}{q}\frac{1}{\sqrt{2\cosh [2x_6]}}\right)V_+^2\right].
\eea


\begin{thebibliography}{99}


\bibitem{'tHooft:1973jz}
  G.~'t Hooft,
  ``A planar diagram theory for strong interactions,''
  Nucl.\ Phys.\  B {\bf 72}, 461 (1974).

\bibitem{AharonyTI}
  O.~Aharony, S.~S.~Gubser, J.~M.~Maldacena, H.~Ooguri and Y.~Oz,
  ``Large N field theories, string theory and gravity,''
  Phys.\ Rept.\  {\bf 323}, 183 (2000)
  [arXiv:hep-th/9905111].


\bibitem{Nunez:2010sf}
  C.~Nunez, A.~Paredes and A.~V.~Ramallo,
  ``Unquenched flavor in the gauge/gravity correspondence,''
  arXiv:1002.1088 [hep-th].




\bibitem{GiveonSR}
  A.~Giveon and D.~Kutasov,
  ``Brane dynamics and gauge theory,''
  Rev.\ Mod.\ Phys.\  {\bf 71}, 983 (1999)
  [arXiv:hep-th/9802067].

\bibitem{Polchinski:2000uf}
  J.~Polchinski and M.~J.~Strassler,
  ``The string dual of a confining four-dimensional gauge theory,''
  arXiv:hep-th/0003136.

\bibitem{ks}
  I.~R.~Klebanov and M.~J.~Strassler,
  ``Supergravity and a confining gauge theory: Duality cascades and chiSB-resolution of naked singularities,''
  JHEP {\bf 0008}, 052 (2000)
  [arXiv:hep-th/0007191].

\bibitem{Maldacena:2000yy}
  J.~M.~Maldacena and C.~Nunez,
  ``Towards the large N limit of pure N = 1 super Yang Mills,''
  Phys.\ Rev.\ Lett.\  {\bf 86}, 588 (2001)
  [arXiv:hep-th/0008001].

\bibitem{Seiberg:1994pq}
  N.~Seiberg,
  ``Electric - magnetic duality in supersymmetric nonAbelian gauge theories,''
  Nucl.\ Phys.\  B {\bf 435}, 129 (1995)
  [arXiv:hep-th/9411149].

\bibitem{Witten:1997ep}
  E.~Witten,
  ``Branes and the dynamics of {QCD},''
  Nucl.\ Phys.\  B {\bf 507}, 658 (1997)
  [arXiv:hep-th/9706109].

\bibitem{ItzhakiDD}
  N.~Itzhaki, J.~M.~Maldacena, J.~Sonnenschein and S.~Yankielowicz,
  ``Supergravity and the large N limit of theories with sixteen
  supercharges,''
  Phys.\ Rev.\  D {\bf 58}, 046004 (1998)
  [arXiv:hep-th/9802042].


\bibitem{Dymarsky:2005xt}
  A.~Dymarsky, I.~R.~Klebanov and N.~Seiberg,
  ``On the moduli space of the cascading $SU(M+p) \times SU(p)$ gauge theory,''
  JHEP {\bf 0601}, 155 (2006)
  [arXiv:hep-th/0511254].

\bibitem{Witten:1997sc}
  E.~Witten,
  ``Solutions of four-dimensional field theories via M-theory,''
  Nucl.\ Phys.\  B {\bf 500}, 3 (1997)
  [arXiv:hep-th/9703166].

\bibitem{Benini:2007gx}
  F.~Benini, F.~Canoura, S.~Cremonesi, C.~Nunez and A.~V.~Ramallo,
  ``Backreacting Flavors in the Klebanov-Strassler Background,''
  JHEP {\bf 0709}, 109 (2007)
  [arXiv:0706.1238 [hep-th]].

\bibitem{AharonyJU}
  O.~Aharony and A.~Hanany,
  ``Branes, superpotentials and superconformal fixed points,''
  Nucl.\ Phys.\  B {\bf 504}, 239 (1997)
  [arXiv:hep-th/9704170].

\bibitem{Klebanov:1998hh}
  I.~R.~Klebanov and E.~Witten,
  ``Superconformal field theory on threebranes at a Calabi-Yau  singularity,''
  Nucl.\ Phys.\  B {\bf 536}, 199 (1998)
  [arXiv:hep-th/9807080].

\bibitem{Klebanov:1999rd}
  I.~R.~Klebanov and N.~A.~Nekrasov,
  ``Gravity duals of fractional branes and logarithmic RG flow,''
  Nucl.\ Phys.\  B {\bf 574}, 263 (2000)
  [arXiv:hep-th/9911096].


\bibitem{Klebanov:2000nc}
  I.~R.~Klebanov and A.~A.~Tseytlin,
  ``Gravity Duals of Supersymmetric $SU(N) \times SU(N+M)$ Gauge Theories,''
  Nucl.\ Phys.\  B {\bf 578}, 123 (2000)
  [arXiv:hep-th/0002159].


\bibitem{Strassler:2005qs}
  M.~J.~Strassler,
  ``The duality cascade,''
  arXiv:hep-th/0505153.

\bibitem{Aharony:2004xn}
  O.~Aharony, A.~Giveon and D.~Kutasov,
  ``LSZ in LST,''
  Nucl.\ Phys.\  B {\bf 691}, 3 (2004)
  [arXiv:hep-th/0404016].

\bibitem{Sakai:2004cn}
  T.~Sakai and S.~Sugimoto,
  ``Low energy hadron physics in holographic QCD,''
  Prog.\ Theor.\ Phys.\  {\bf 113}, 843 (2005)
  [arXiv:hep-th/0412141].

\bibitem{Antonyan:2006vw}
  E.~Antonyan, J.~A.~Harvey, S.~Jensen and D.~Kutasov,
  ``NJL and QCD from string theory,''
  arXiv:hep-th/0604017.

\bibitem{VanRaamsdonk:2009gh}
  M.~Van Raamsdonk and K.~Whyte,
  ``Baryons from embedding topology and a continuous meson spectrum in a new
  holographic gauge theory,''
  arXiv:0912.0752 [hep-th].

\bibitem{Pasti:1997gx}
  P.~Pasti, D.~P.~Sorokin and M.~Tonin,
  ``Covariant action for a D = 11 five-brane with the chiral field,''
  Phys.\ Lett.\  B {\bf 398} (1997) 41
  [arXiv:hep-th/9701037].


\bibitem{Brandhuber:1998er}
  A.~Brandhuber, N.~Itzhaki, J.~Sonnenschein and S.~Yankielowicz,
  ``Wilson loops, confinement, and phase transitions in large N gauge  theories
  from supergravity,''
  JHEP {\bf 9806}, 001 (1998)
  [arXiv:hep-th/9803263].

\bibitem{Witten:1998zw}
  E.~Witten,
  ``Anti-de Sitter space, thermal phase transition, and confinement in  gauge
  theories,''
  Adv.\ Theor.\ Math.\ Phys.\  {\bf 2}, 505 (1998)
  [arXiv:hep-th/9803131].

\bibitem{Aharony:2006da}
  O.~Aharony, J.~Sonnenschein and S.~Yankielowicz,
  ``A holographic model of deconfinement and chiral symmetry restoration,''
  Annals Phys.\  {\bf 322}, 1420 (2007)
  [arXiv:hep-th/0604161].

\bibitem{Parnachev:2006dn}
  A.~Parnachev and D.~A.~Sahakyan,
  ``Chiral phase transition from string theory,''
  Phys.\ Rev.\ Lett.\  {\bf 97}, 111601 (2006)
  [arXiv:hep-th/0604173].

\bibitem{Antonyan:2006qy}
  E.~Antonyan, J.~A.~Harvey and D.~Kutasov,
  ``The Gross-Neveu model from string theory,''
  Nucl.\ Phys.\  B {\bf 776}, 93 (2007)
  [arXiv:hep-th/0608149].

\end{thebibliography}
\end{document}